\documentclass{rQUF2e}

\usepackage{graphicx} 
\usepackage{subfigure}

\usepackage{natbib}

\usepackage{algorithm}
\usepackage{algorithmic}

\usepackage{hyperref}
\usepackage{microtype}      
\usepackage{relsize}

\usepackage{tikz}

\usepackage{amsmath}
\usepackage{amssymb}
\usepackage{bbm}
\usepackage{booktabs}

\newcommand{\bG}{\boldsymbol{G}}

\newcommand{\bC}{\boldsymbol{C}}

\newcommand{\blamt}{\boldsymbol{\lambda_t}}
\newcommand{\bK}{\textbf{\textsf{K}}}
\newcommand{\bI}{\boldsymbol{I_d}}

\newcommand{\bLam}{\boldsymbol{\Lambda}}

\newcommand{\bR}{\boldsymbol{R}}

\newcommand{\bN}{\boldsymbol{N}}

\newcommand{\bKc}{\boldsymbol{K^c}}
\newcommand{\bmu}{\boldsymbol{\mu}}

\newcommand{\E}{\mathbb{E}}

\newcommand{\bPsi}{\boldsymbol{\Psi}}

\title{{Analysis of order book flows using a nonparametric estimation of the branching ratio matrix}}
\author{M. Achab$\dag$, E. Bacry$\dag$, J. F. Muzy$\dag,\ddag$ and M. Rambaldi$\dag$ \\}
\affil{$\dag$CMAP, Ecole Polytechnique, CNRS, UMR 7641, 91128 Palaiseau, France \\
$\ddag$SPE, Universit\'e de Corse, CNRS, UMR 6134, Campus Grimaldi, 20250 Corte, France}

\begin{document}

\maketitle

\begin{abstract}
We introduce a new non parametric method that allows for a direct, fast and efficient estimation of the matrix of kernel norms of a multivariate Hawkes process, also called branching ratio matrix.
We demonstrate the capabilities of this method by applying it to high-frequency order book data from the EUREX exchange. We show that it is able to uncover (or recover) various relationships between all the first level order book events associated with some asset when mapped to a 12-dimensional process. We then scale up the model so as to account for events on two assets simultaneously and we discuss the joint high-frequency dynamics.
\end{abstract}



\begin{classcode}C14, C58. \end{classcode}

\section{Introduction}
With the large number of empirical studies devoted to high frequency finance, relying on
datasets of increasing size and quality, many progresses have been made during
the last decade in the modelling and understanding the microstructure of financial markets.
Within this context, as evidenced by this special issue, Hawkes processes
have become a very popular class of models. The main reason is that they allow one
to account for the mutual influence of various types of events
in a simple and parsimonious way through a conditional intensity vector.
Hawkes processes have been involved in many different problems of high frequency finance ranging from
the simple description of the temporal occurrence of market orders or price changes (\cite{Bowsher2007, HardimanBouchaud2014, filimonov2012quantifying}),  to the complex
modelling of the arrival rates of various kinds of events in a full order book model (\cite{large2007measuring, toke2011market, jedidi2013stability}). We refer to \cite{HawkesFinanceReview} for a recent review.

A multivariate Hawkes model of dimension $d$ is characterized by a $d \!\times \! d$ matrix of kernels, whose elements $\phi^{ij}(t)$ account for the influence, after a lag $t$, of events of type $j$ on the arrival rate of events of type $i$. The challenging issue of the statistical estimation of the shape of these excitation kernels has been addressed by many authors
and various solutions have been proposed whose performances (accuracy and computational complexity) strongly depend on the empirical situation one considers. Indeed, if non-parametric methods like e.g. the EM method
(\cite{Lewis:2011aa}), the Wiener-Hopf method (\cite{Bacry:2014aa,bacry14,thibault}) or the contrast function method (\cite{reynaud2014goodness}) can be applied in low dimensional situations with a large number of events,
one has to consider parametric penalized alternatives (like e.g., in \cite{zhou_zha_le_2013,yang2013mixture}) when one has to handle a system of very large dimension with a relative low number of observed events (as, e.g., when studying events
associated with the node activities of some social networks).

As far as (ultra) high frequency finance is concerned, the overall number of events can be very large.
These events occur in a very correlated manner (with long-range
correlations) and the system dimensionality can vary from low to moderately high.
In a series of recent papers, Bacry {\em et al.} have shown
that the non parametric Wiener-Hopf method provides reliable estimations in order to describe, within
a multivariate Hawkes model, various aspects of level-I order book fluctuations: the coupled
dynamics of mid-price changes, market and limit order arrivals (\cite{Bacry:2014aa,thibault}), the impact of market orders (\cite{bacry2015impact}) or the interplay between book orders of different sizes  (\cite{rambaldi2016role}). However, if one wants to account for systems of larger dimensionality by
considering for instance a wider class of event types or the book events associated with a basket
(e.g. a couple) of assets, then the Wiener-Hopf method (or any other similar non-parametric method) may reach
its limits as respect to both computational cost and estimation accuracy. On the other hand, a parametric approach
can lead to strong bias in the estimated influences between components.

For this reason, in the present paper, we propose to estimate Hawkes models of order book data using
the faster and simpler non-parametric
approach introduced in \cite{achab2016uncovering}. This method
focuses only on the global properties of the Hawkes process. More precisely, it aims at estimating directly the matrix of the kernel norms (also called the {\em branching ratio matrix}) without paying attention to the precise shape of these kernels. As recalled in the next section, this matrix does not bring
all the information about the process dynamics, but is sufficient to disentangle the complex interactions between various type of events and estimate
the magnitude of their self- and cross- excitations.
Moreover, it allows one to estimate the amplitude of fluctuations of endogenous origin
as compared to those of exogenous sources.
The method we propose can be considered as the multivariate extension of the approach pioneered
by \cite{HardimanBouchaud2014} that proposed to estimate the kernel norm of a one-dimensional Hawkes model directly from the integral of the empirical correlation function.
Unfortunately their approach cannot be immediately extended to a multivariate framework because it does not bring a sufficient number of constraints as compared to the number of unknown parameters.
The method of  \cite{achab2016uncovering} circumvents this difficulty by taking into account
the first three integrated cumulant tensors of Hawkes process.

The paper is organized as follows: in Section \ref{sec:Hawkes} we provide the main definitions and properties
of multivariate Hawkes processes and we introduce the main notations we use all along the paper.
The cumulant method of Achab et al. is described and illustrated in Section \ref{sec:nphc}.
In Section \ref{sec:emp} we estimate the matrix of kernel of Hawkes models for level-I book events associated
with 4 differents very liquid assets, namely DAX, Euro-Stoxx, Bund and Bobl future contracts.
We first consider the 8-dimensional model proposed in \cite{thibault} in order to compare our method to the former results obtained with a computationally more complex Wiener-Hopf method. We then show that the cumulant approach can easily be extended to a 12-dimensional model where all types of level-I book events are considered. Within this model, we uncover all the relationships
between these types of events and we study the daily amplitude variations of exogenous intensities.
In Section \ref{sec:multi} we investigate the correlation between two assets
by considering the events of their order book within a 16-dimensional model. This allows us to discuss the influence of both their tick size and their degree of reactivity with respect to the impact of their book events on each other.
Section \ref{sec:conclusion} contains concluding remarks while
some technical details are provided in Appendix.

\section{Hawkes processes: definitions and properties}\label{sec:Hawkes}

In this section we provide the main definitions and properties of multivariate Hawkes processes
and set the notations we need all along the paper.

\subsection{Multivariate Hawkes processes and the branching ratio matrix $\bG$}
A multivariate Hawkes process of dimension $d$ is a $d$-dimensional counting processes $\bN_t$ with
a conditional intensity vector $\blamt$ that is a linear function of past events. More precisely,
\begin{equation}
\label{eq:defHawkes}
\lambda^i_t = \mu^i+ \sum_{j=1}^d \int_{-\infty}^t \phi^{ij}(t-s) \; dN^j_s
\end{equation}
where $\mu^i$ represents the baseline intensity while the kernel $\phi^{ij}(t)$ quantifies the excitation rate of an event of type $j$ on the arrival rate of events of type $i$ after a time lag $t$. In general it is assumed that each kernel is causal and positive, meaning that Hawkes processes can only
account for mutual excitation effects since the occurrence of some event can only
increase the future arrival intensity of other events. In order to consider the possibility of
inhibition effects, one can allow kernels to take negative values. In that case, we have to consider
expression \eqref{eq:defHawkes} only when it provides a positive result while the conditional intensity
is assumed to be zero otherwise.
Rigorously speaking, such non-linear variant of Eq. \eqref{eq:defHawkes} cannot be handled as simply as the original Hawkes process (\cite{bremaud1996stability}) but, as empirically shown in e.g. \cite{reynaud2014goodness} or \cite{bacry14}, if the probability that
$\lambda^i_t < 0$ is small enough, one can safely consider the model as linear so that all standard expressions provide accurate results.
In the following we will suppose that we are in this case and
we don't necessarily impose that the kernels $\phi^{ij}(t)$ are positive functions.

Let us define the matrix $\bG$ as the matrix whose coefficients are the integrals of the
kernels $\phi^{ij}(t)$ (that are supported by ${\rm I\!R}^+$):
\begin{equation}
\label{eq:defG}
G^{ij} = \int_0^{+\infty} \phi^{ij}(t) dt \; .
\end{equation}
Let us remark that, as it can directly be seen from
the cluster representation of Hawkes processes (\cite{hawkes74}), $G^{ij}$ represents the mean total
number of events of type $i$ directly triggered by an event of type $j$. For that reason, in the literature,  the matrix
$\bG$ is also referred to as the {\em branching ratio matrix}  (\cite{HardimanBouchaud2014}).
Notice that since the kernels $\phi^{ij}(t)$ are not necessarily non negative functions, $G^{ij}$
does not in general correspond to the $L^1$ norm of $\phi^{ij}$.
For the sake of simplicity, though this is not technically correct,
we shall often refer to the matrix $\bG$ as the ``matrix of kernel norms" or more simply the ``norm matrix".

If $\Vert \bG \Vert$ stands for the largest eigenvalue of $\bG$,
it is well known that a sufficient condition for the intensity process $\blamt$ to be stationary is that $\Vert \bG \Vert < 1$. In the following we will always consider this condition satisfied. 
One can then define the matrix $\bR$ as:
\begin{equation}
\label{eq:defR}
\bR = (\bI-\bG)^{-1},
\end{equation}
where $\bI$ denotes the identity matrix of dimension $d$.

Let $\bLam$ denote the mean intensity vector:
\begin{equation}
\label{eq:defLambda}
\bLam= \E(\blamt) \; ,
\end{equation}
so that the ratio $\frac{\mu^{i}}{\Lambda^i}$ represents the fraction of events of type $i$ that are of exogenous
origin.
One can easily prove that $\bLam$ and $\bmu$ are related as:
\begin{equation}
\label{eq:rel_mu_lambda}
\bLam = \bR \; \bmu
\end{equation}

If one defines the matrix $\bPsi$ as:
\begin{equation}\label{eq:defpsi}
\bPsi = \bG \bR \; = \bR - \bI,
\end{equation}
then $\Psi^{ij}$ represents the average number of events of type $i$ triggered (directly or indirectly) by an exogenous
event of type $j$.
When one analyzes empirical data within the framework of Hawkes processes, the previous remarks allow one
to quantify causal relationships between events in the sense of Granger, i.e., within a well defined
mathematical model. In that respect, the coefficients of the matrices
$\bG$ or $\bPsi$ can be read as (Granger-)causality relationships between various types of events and used as a tool to disentangle the complexity of the observed flow of events occurring in some experimental situations (\cite{eichler2017graphical}).
Let us emphasize that such causal implications are just a matter of interpretation of data within a specific model
(namely a Hawkes model) and should simply be considered as a convenient and parsimonious way to represent that data. They
should not, in any way, be understood as a ``physical" causality reflecting their ``real nature''.

\subsection{Integrated Cumulants of Hawkes Process}
The NPHC algorithm developed in \cite{achab2016uncovering} and described in Sec. \ref{sec:nphc} below,
enables the direct estimation of the matrix $\bG$ from a single or several realizations of the process. It relies on the computation
of low order cumulant functions whose expressions are recalled below.

Given $1 \leq i,j,k \leq d$, the first three integrated cumulants of the Hawkes process can be, thanks to stationarity,
defined as follows:
\begin{align}
\label{eq:cumul1density}
\Lambda^i dt &= \mathbb{E}(dN_t^i) \\
\label{eq:cumul2density}
C^{ij} dt &= \int_{\tau \mathsmaller{\in} \mathbb{R}} \! \Big(
\mathbb{E}(dN_{t}^{i} dN_{t+\tau}^{j}) - \mathbb{E}(dN_{t}^{i}) \mathbb{E}(dN_{t+\tau}^{j}) \Big) \\
K^{ijk} dt &= \int \! \! \int_{\tau,\tau' \mathsmaller{\in} \mathbb{R}^2} \! \Big(
\mathbb{E}(dN^i_t dN^j_{t+\tau} dN^k_{t+\tau'}) + 2 \mathbb{E}(dN^i_t) \mathbb{E}(dN^j_{t+\tau}) \mathbb{E}(dN^k_{t+\tau'}) \nonumber \\
& \quad \quad \quad - \mathbb{E}(dN^i_t dN^j_{t+\tau}) \mathbb{E}(dN^k_{t+\tau'}) - \mathbb{E}(dN^i_t dN^k_{t+\tau'}) \mathbb{E}(dN^j_{t+\tau}) - \mathbb{E}(dN^j_{t+\tau} dN^k_{t+\tau'}) \mathbb{E}(dN^i_t)
\Big),
\label{eq:cumul3density}
\end{align}
where Eq.~(\ref{eq:cumul1density}) is the mean intensity of the Hawkes process, the second-order cumulant~(\ref{eq:cumul2density}) refers to the integrated covariance density matrix and the third-order cumulant~(\ref{eq:cumul3density}) measures the skewness of~$\boldsymbol{N}_t$.
Using the martingale representation~(\cite{bacry14}) or the Poisson cluster process representation~(\cite{jovanovic2015cumulants}), one can obtain an explicit relationship between these integrated cumulants and the matrix
$\boldsymbol{R}$ (and therefore the matrix $\bG$ thanks to Eq. \eqref{eq:defR}).
Some straightforward computations (see \cite{achab2016uncovering}) lead to the following identities:
\begin{align}
\label{eq:relation0}
\Lambda^i  &=  \sum_{m=1}^d R^{im} \mu^m  \\
\label{eq:relation1}
C^{ij} &=  \sum_{m=1}^d \Lambda^m R^{im} R^{jm} \\
\label{eq:relation2}
K^{ijk}  &=  \sum_{m=1}^d ( R^{im}R^{jm}C^{km} + R^{im}C^{jm}R^{km} + C^{im}R^{jm}R^{km} - 2 \Lambda^m R^{im}R^{jm}R^{km}).
\end{align}

\section{The NPHC method}\label{sec:nphc}
In this section we briefly recall the main lines of the recent non parametric method proposed
in \cite{achab2016uncovering}
that leads to a fast and robust direct estimation of the branching ratio matrix $\bG$ without
estimating the shape of the kernel functions. This method is based on the remark that, as
shown in \cite{jovanovic2015cumulants} and as it can be seen in Eqs. \eqref{eq:relation0}, \eqref{eq:relation1}
and \eqref{eq:relation2}, the integrated cumulants of a Hawkes process
can be explicitly written as functions of $\bR$. The NPHC method is
a moment method that consists in directly exploiting these equations to recover $\bR$ and thus $\bG$.

\subsection{Estimation of the integrated cumulants}
Let us first introduce explicit formulas to estimate the three moment-based quantities listed in the previous section, namely, $\bLam$, $\bC$ and $\bK$.
In what follows, we assume there exists $H>0$ such that the truncation from $(-\infty,+\infty)$ to $[-H,H]$ of the domain of integration of the quantities appearing in Eqs.~(\ref{eq:cumul2density}) and~(\ref{eq:cumul3density}) introduces only a small error.
This amounts to neglecting tail effects in the covariance density and in the skewness density, and it corresponds to a good approximation if $(i)$~each kernel $\phi^{ij}(t)$ is essentially supported by $[0,H]$ and $(ii)$ the spectral norm $\Vert \bG \Vert$ is less than $1$.

\noindent
In this case, given a realization of a stationary Hawkes process $\{ \boldsymbol N_t : t \in [0, T] \}$, as shown in \cite{achab2016uncovering}, we can write the estimators of the first three cumulants~(\ref{eq:cumul1density}),~(\ref{eq:cumul2density}) and~(\ref{eq:cumul3density}) as
\begin{align}
    \label{eq:estimator1}
    \widehat \Lambda^i &= \frac{1}{T} \sum_{\tau \in Z^i} 1 = \frac{N^i_T}{T} \, \\
    \label{eq:estimator2}
    \widehat{C}^{ij} &= \frac{1}{T} \sum_{\tau \in Z^i} \left( N^j_{\tau+H} - N^j_{\tau-H} - 2 H \widehat{\Lambda}^j \right) \, \\
    \begin{split}
    \widehat{K}^{ijk} &= \frac{1}{T} \sum_{\tau \in Z^i} \left( N^j_{\tau+H} - N^j_{\tau-H} - 2 H \widehat{\Lambda}^j \right)  \cdot\left( N^k_{\tau+H} - N^k_{\tau-H} - 2 H \widehat{\Lambda}^k \right) \\
    & \quad - \frac{\widehat \Lambda^i}{T} \sum_{\tau \in Z^j} \sum_{\tau' \in Z^k} (2 H - | \tau'-\tau|)^{+} + 4 H^2 \widehat{\Lambda}^i \widehat{\Lambda}^j \widehat{\Lambda}^k.
    \end{split}
    \label{eq:estimator3}
\end{align}

In practice, the filtering parameter $H$ is selected by $(i)$ computing estimates of the covariance density at several points $t$ \footnote{the pointwise covariance density at $t$ can be estimated with $\frac{1}{h T} \sum_{\tau \in Z^i} \left( N^j_{\tau+t+h} - N^j_{\tau+t} - h \widehat{\Lambda}^j \right)$ for a small $h$},
$(ii)$ assessing the characteristic time $\tau_c$ after which the covariance density is negligible, and $(iii)$ setting a multiple of $\tau_c$ for $H$, for instance $H = 5 \tau_c$.

\subsection{The NPHC algorithm}
The covariance $\bC$ only provides $d(d+1)/2$ independent coefficients and is therefore not sufficient to uniquely identify the $d^2$ coefficients of the matrix $\bG$.
In order to set a sufficient number of constraints, the NPHC approach relies on using all the covariance $\bC$ along with a restricted number of the $(d^3+3d^2+2d)/6$ third-order independent cumulant components, namely the $d^2$ coefficients
$\bKc= \{ K^{iij}\}_{1\leq i,j \leq d}$.
Thus, we define the estimator of $\bR$ as $\widehat \bR \in \textrm{argmin}_{\bR} \mathcal{L}(\bR)$, where
\begin{align}
	\label{eq:nphc_loss}
  &\mathcal{L}(\bR) = (1 - \kappa) \|\bKc(\bR) - \widehat{\bKc}\|_2^2 + \kappa \| \bC (\bR) - \widehat{\bC} \|_2^2,
\end{align}
where $\| \cdot \|_2$ stands for the Frobenius norm,
while $\widehat{\bKc}$ and $\widehat{\bC}$ are the respective estimators of $\bC$ and $\bKc$ as defined in Equations~\eqref{eq:estimator2}, \eqref{eq:estimator3} above.
It is noteworthy that the above mean square error approach can be seen as a peculiar instance of Generalized Method of Moments (GMM), see~\cite{hall}, \cite{hansen1982large}.
Though this framework allows to determine the optimal weighting matrix involved in the loss function, in practice this approach is unusable, as the associated complexity is too high. Indeed,
since we have $d^2$ parameters, this matrix has $d^4$ coefficients and GMM calls for computing its inverse leading to a $O(d^6)$ complexity.
Thus, instead, we choose to use the loss function \eqref{eq:nphc_loss} in which, so as to be of the same order, the two terms are rescaled using  $\kappa = \|\widehat{\bKc}\|^2_2 / (\|\widehat{\bKc}\|^2_2 +
  \|\widehat{\bC}\|^2_2)$.
We refer to Appendix \ref{appen:origin_kappa} for an explanation of how $\kappa$ is related to the weighting matrix.
Finally the estimator of $\bG$ is straightforwardly obtained as
\begin{equation*}
	\widehat{\bG} = \bI - \widehat{\bR}^{-1},
\end{equation*}
from the inversion of Eq.~\eqref{eq:defG}.
The authors of \cite{achab2016uncovering} proved the consistency of the so-obtained estimator $\widehat{\bG}$, i.e. the convergence in probability to the true value, when the observation time $T$ goes to infinity.

Let us mention that, when applied to financial time-series, the number of events is generally large as compared with $d$ (i.e., $n = \max_i |Z^i| \gg d$), thus
the matrix inversion in the previous formula is not the bottleneck of the algorithm. Indeed, it has a complexity $O(d^3)$ which is cheap as compared with the computation of the cumulants which is $O (n d^2)$.
Thus, assuming the loss function (\ref{eq:nphc_loss}) is minimized after $N_{\text{iter}}$ iterations, the overall complexity of the algorithm is $O (n d^2 + N_{\text{iter}} d^3 )$.
The authors of \cite{achab2016uncovering} compared the complexity of their algorithm with other state-of-the-art methods' ones,
namely the ordinary differential equations based (ODE) algorithm in~\cite{zhou2013learning}, the Sum of Gaussians based algorithm in \cite{xu2016learning},
the ADM4 algorithm in \cite{zhou_zha_le_2013}, and the Wiener-Hopf-based algorithm in \cite{bacry14}.
The complexity of NPHC is smaller, because the algorithm NPHC directly estimates the kernels' integrals while other methods go through the estimation of the kernel functions themselves.

\subsection{Numerical experiments}
\label{subsec:num}

\begin{figure}
\subfigure[Rectangular kernel]
{
\begin{tikzpicture}[scale=1.10]
\draw[->] (0,0) -- (6,0) node[anchor=north] {$t$};
\draw[->] (0,0) -- (0,3) node[anchor=east] {$\phi_t$};
\draw	(0,0) node[anchor=north] {0}
		(1,0) node[anchor=north] {$\gamma$}
		(3,0) node[anchor=north] {$\gamma + 1/\beta$}
        (0,2) node[anchor=east] {$\alpha \beta$};
\draw[thick] (0,0) -- (1,0) -- (1,2) -- (3,2) -- (3,0) -- (6,0);
\draw[thick,dashed] (0,2) -- (1,2);
\end{tikzpicture}
}
\subfigure[Power-law kernel on log-log scale]
{
\begin{tikzpicture}[scale=1.10]
\draw[->] (0,0) -- (6,0) node[anchor=north] {$\log t$};
\draw[->] (0,0) -- (0,3) node[anchor=east] {$\log \phi_t$};
\draw	(0,0) node[anchor=north] {}
		(2,0) node[anchor=north] {$- \log \beta$}
		(0,2) node[anchor=east] {$\log \alpha \beta \gamma$};
\draw (4.7,0.8) node{{\scriptsize slope $\approx -(1+\gamma$)}};
\draw[thick] [domain=0:4.1] plot(\x, {2-0.5*ln(1+10^(\x-2))});
\draw[thick,dashed] (2,0) -- (2,1.6);
\end{tikzpicture}
}
\caption{The two different kernels used to simulate the datasets. }
\label{fig:kernels}
\end{figure}
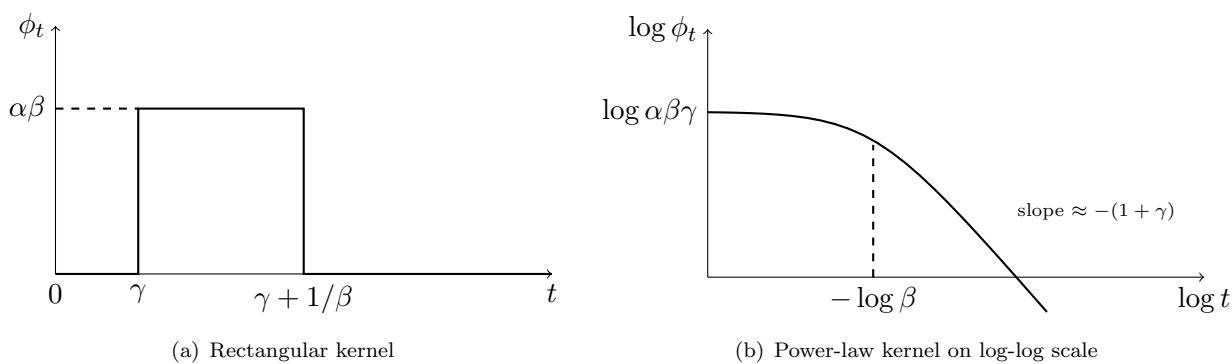

As mentioned above, the NPHC algorithm is non parametric and provides an estimation
of the integral of the kernels regardless of their shapes.
In order to illustrate the stability of our method with respect to the shape of the kernels, we simulated two datasets
with Ogata's Thinning algorithm introduced in \cite{ogata1981lewis} using the open-source library \texttt{tick}\footnote{\url{https://github.com/X-DataInitiative/tick}}. Each dataset
corresponds to a different kernel shape (but with the same norm),  a rectangular kernel and a power-law kernel, both represented in Figure \ref{fig:kernels}:
\begin{align}
\label{eq:rect_kernel}
&\mbox{rectangular kernel: }\phi (t) = \alpha \beta \mathbbm{1}_{[0, 1 / \beta]}(t - \gamma) \\
\label{eq:plaw_kernel}
&\mbox{power law kernel: } \mbox{ }\phi (t) = \alpha \beta \gamma (1 + \beta t)^{-(1 + \gamma)}
\end{align}

\begin{figure}[tbh]
\label{fig:nphc_simu}
\centering
\includegraphics[width=\textwidth]{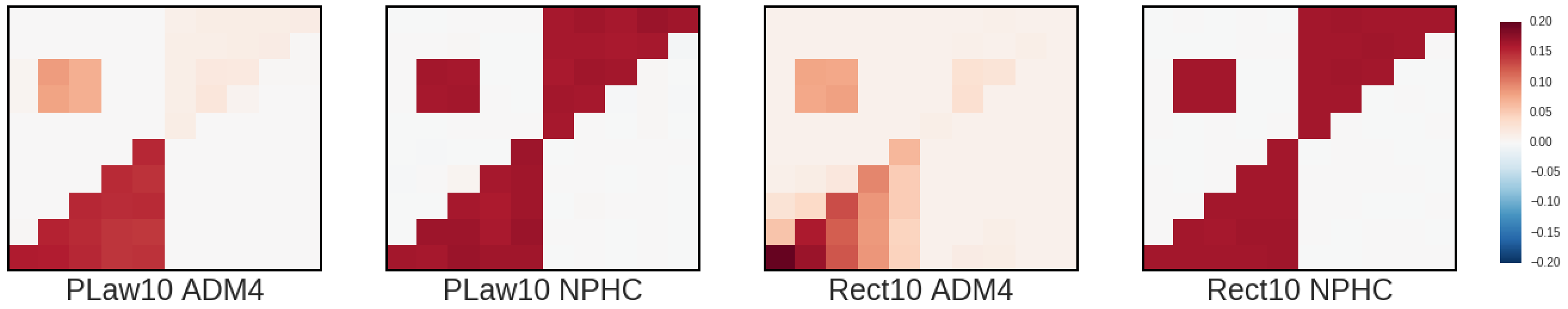}
\caption{Estimated matrices $\bG$ with our NPHC algorithm and the ADM4 algorithm  from \cite{zhou_zha_le_2013} on the two datasets Rect10 and PLaw10. NPHC shows significantly better results on these two datasets.}
\end{figure}

In both cases, $\alpha$ corresponds the integral of the kernel, $1/\beta$ can be regarded as a characteristic time-scale, and $\gamma$ corresponds to the scaling exponent for the power law kernel and a delay parameter for the rectangular one.
We consider a non-symmetric 10-dimensional block-matrix $\bG$ with 3 non-zero blocks, and where the parameters $\alpha = 1 / 6$ and $\gamma = 1/2$ take the same constant values on these blocks.
Three different $\beta_0$, $\beta_1$ and $\beta_2$ are used in the different blocks, with $\beta_{2} / \beta_1 = \beta_{1} / \beta_0 = 10$ and $\beta_0 = 0.1$.
The number of events is roughly equal to $10^5$ on average over the nodes. We thus obtain
two datasets, the first one referred to as Rect10 corresponding to the rectangular kernels and the second one referred to as PLaw10 corresponding to the power law kernels. We run on these two datsets the NPHC algorithm and the ADM4 algorithm from \cite{zhou_zha_le_2013}, which calibrates
a single exponential kernel $t \rightarrow \alpha \beta e^{-\beta t}$ with constant $\beta$, and for which we provided the intermediate true value $\beta = \beta_1$.
The results are shown in Figure \ref{fig:nphc_simu}.
These figures clearly illustrate that parametric methods can lead to very poor results when the parametrization does not represent well the data,
while NPHC method gives better solutions without knowing scaling parameters $\beta$.

\section{Single-asset model}\label{sec:emp}
In this section we apply the NPHC method to high-frequency financial data. First we describe our dataset,
then we compare the results of the NPHC method with those obtained with the Wiener-Hopf method of \cite{bacry14} on the 8-dimensional model of single asset level-I book order events proposed in \cite{thibault}. We finally discuss the NPHC estimation of the
norm matrix associated with a ``complete version'' (i.e. 12-dimensional) of this model.

\subsection{Data}\label{sec:data}
In this paper we use level-I order book data provided by QuantHouse EUROPE/ASIA
(http://www.quanthouse.com) for four future contracts traded on the Eurex exchange, namely the futures on the DAX and Euro Stoxx 50 equity indices, and the Bund and Bobl futures. The DAX and Euro Stoxx 50 indices track the largest stock by market capitalization in Germany and the Euro area respectively, while the Bund and Bobl are German interest rate futures on the 8.5 -10.5 years and the 4.5-5.5 years horizon respectively.
The data span a period of 338 trading days from July 2013 to October 2014. For each asset, a line with the current status of the first levels of the order book is added to the database every time there is a change (price, volume or both). Moreover, an additional line is added in the case the change is caused by a market order and a trade is generated. It is therefore possible to obtain a list of the orders that were submitted complete with their time, type (limit, cancel or market order), volume and price. The timestamp precision is one microsecond and the timestamps are set directly by the exchange.

In this work we are interested in disentangling the interactions of different types of events occurring at the first level of the order book. To this end, we will distinguish the following event types:
\begin{itemize}
\item $T^+$ ($T^-$) : upwards (downwards) mid price movement triggered by a market order;
\item $L^+$ ($L^-$) : upwards (downwards) mid price movement triggered by a limit order;
\item $C^+$ ($C^-$) : upwards (downwards) mid price movement triggered by a cancel order;
\item $T^a$ ($T^b$) : market order at the ask (bid) that does not move the mid price;
\item $L^a$ ($L^b$) : limit order at the ask (bid) that does not move the mid price;
\item $C^a$ ($C^b$) : cancellation order at the ask (bid) that does not move the mid price.
\end{itemize}
Additionally, we introduce the symbols $P^+$ ($P^-$) to denote an upwards (downwards) mid price movement irrespectively of its origin.  In Table~\ref{tab:events} we report the average number of events per day (from 08:00 am to 10:00 pm) for each asset and each type. We remark that all four assets are extremely active securities with an average of more than 300.000 events per day.

One characteristic that strongly influences the order book dynamics at short time scales is the tick size to average spread ratio. When this ratio is close to one (resp. much smaller than one), the asset is said to be a ``large tick asset'' (resp. a ``small tick asset'') (see, e.g., \cite{dayri2015large}). In our dataset, all assets are large-tick assets (the spread is equal to one tick in more than 95\% of the times) except for the DAX future, which is a small-tick one. As evidenced by Table~\ref{tab:events}, the price changes much less frequently on large tick assets. One can also remark that the quantity available at the best quotes tends do be proportionally much larger on large tick assets. These microstructural characteristics will be reflected by our analysis.

\begin{table}
\centering
\begin{tabular}{lrrrrrrrrrrrr}
\toprule
{} &  $T^+$ &  $T^-$ &  $L^+$ &  $L^-$ &  $C^+$ &  $C^-$ &  $T^a$ &  $T^b$ &  $L^a$ &  $L^b$ &  $C^a$ &  $C^b$ \\
\midrule
DAX &   11.9 &   11.9 &   21.8 &   21.9 &   10.1 &   10.1 &   11.6 &   11.7 &   80.0 &   79.5 &   97.3 &   96.1 \\
ESXX &    2.6 &    2.6 &    3.5 &    3.6 &    0.9 &    0.9 &   16.4 &   16.5 &  176.0 &  174.7 &  172.4 &  170.8 \\
Bund &    3.2 &    3.2 &    4.0 &    4.0 &    0.8 &    0.8 &   14.5 &   14.7 &  125.4 &  125.0 &  111.5 &  110.7 \\
Bobl &    1.1 &    1.1 &    1.5 &    1.5 &    0.5 &    0.5 &    6.1 &    6.1 &   86.5 &   86.8 &   81.6 &   81.4 \\
\bottomrule
\end{tabular}
\caption{Average number of events in thousands per type in a trading day (from open at 08:00 to closing at 22:00 Frankfurt time) for the four assets considered.}
\label{tab:events}
\end{table}

%

\subsection{Revising the 8-dimensional mono-asset model of \cite{thibault} : A sanity check}\label{sec:compwh}

In \cite{Bacry:2014aa,bacry14}, the authors outlined a method for non-parametric estimation of the Hawkes kernel functions based the infinitesimal covariance density and the numerical solution of a Wiener-Hopf system of integral equations that links the covariance matrix and the kernel matrix. Their method has been applied to high-frequency financial data in \cite{Bacry:2014aa,thibault}, and \cite{rambaldi2016role}.

The aim of this section is to compare the newly proposed NPHC methodology with the Wiener-Hopf method mentioned above in order to assess the reliability of the new NPHC method. To this end, we reproduce the results obtained in \cite{thibault}.

As it was done there, we consider the DAX and Bund futures data\footnote{Note that we use the very same dataset as in \cite{thibault}} and for each asset we separate Level-I order book events into 8 categories as defined above: $P^+$, $P^-$, $T^a$, $T^b$, $L^a$, $L^b$, and  $C^a$, $C^b$.
Note that here a price move can be of any type. We then consider the timestamp associated with all events as a realization of a 8-dimensional Hawkes process and we use both the NPHC method outlined in Section~\ref{sec:nphc} and the Wiener-Hopf method of \cite{bacry14} to estimate the integrated kernel interaction matrix $\bG$ from the data. For the Wiener-Hopf method, we follow the same procedure as \cite{thibault} and in particular we estimate the covariance density up to a maximum lag of $\approx 1000s$ using a log-linear spaced grid\footnote{As was done in \cite{thibault}, for the estimation of the covariance density we take a linearly spaced grid at short time lags (until a lag of $1ms$) and we switch to a log-spaced one for longer time lags. This allows to estimate the covariance on several orders of magnitude in time.}, while for the NPHC method we follow the steps outlined in Section~\ref{sec:nphc} and we fix $H=500s$ so to be on a comparable scale with the Wiener-Hopf method. Let us note that this scale is several orders of magnitude larger than the typical inter-event time.  Indeed, on the assets considered median inter-event times are of the order of $300\mu s$ (the mean being $\approx 50 ms$), with minimum time distances in the tens of microseconds.

\begin{figure}
\centering
\includegraphics[width=.48\textwidth]{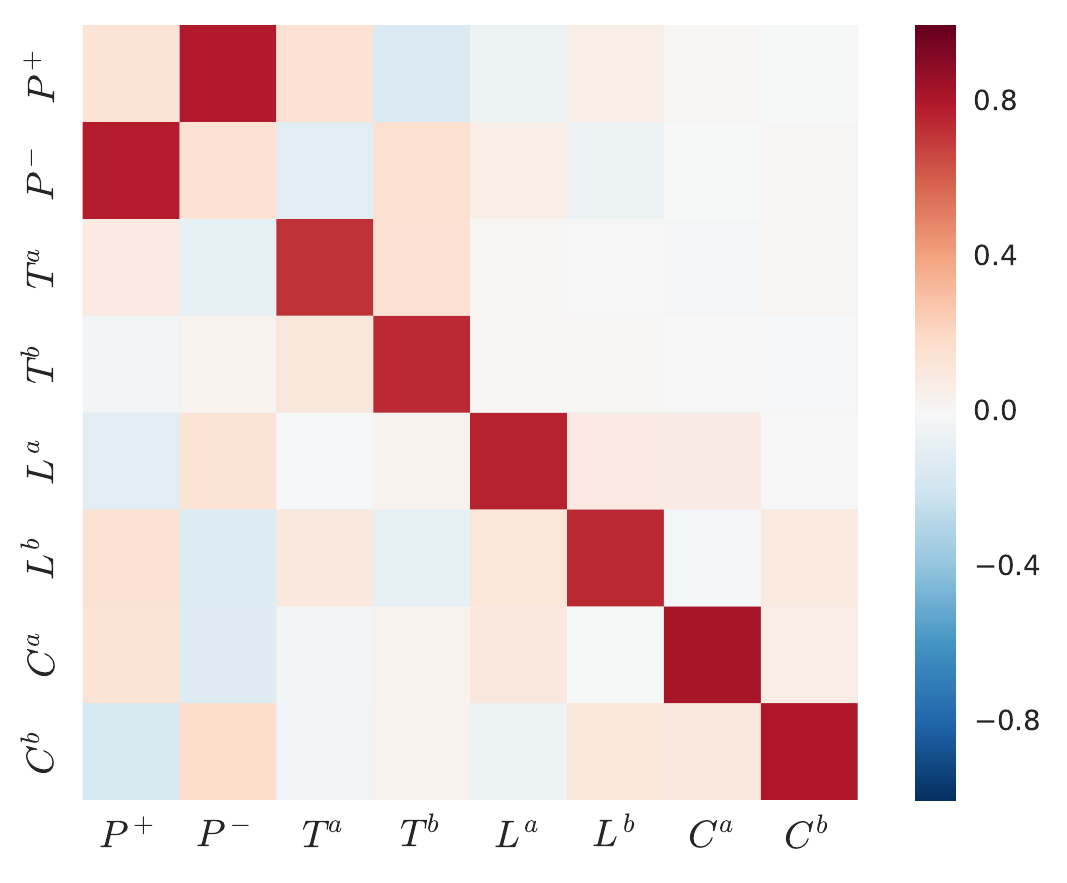}
\includegraphics[width=.48\textwidth]{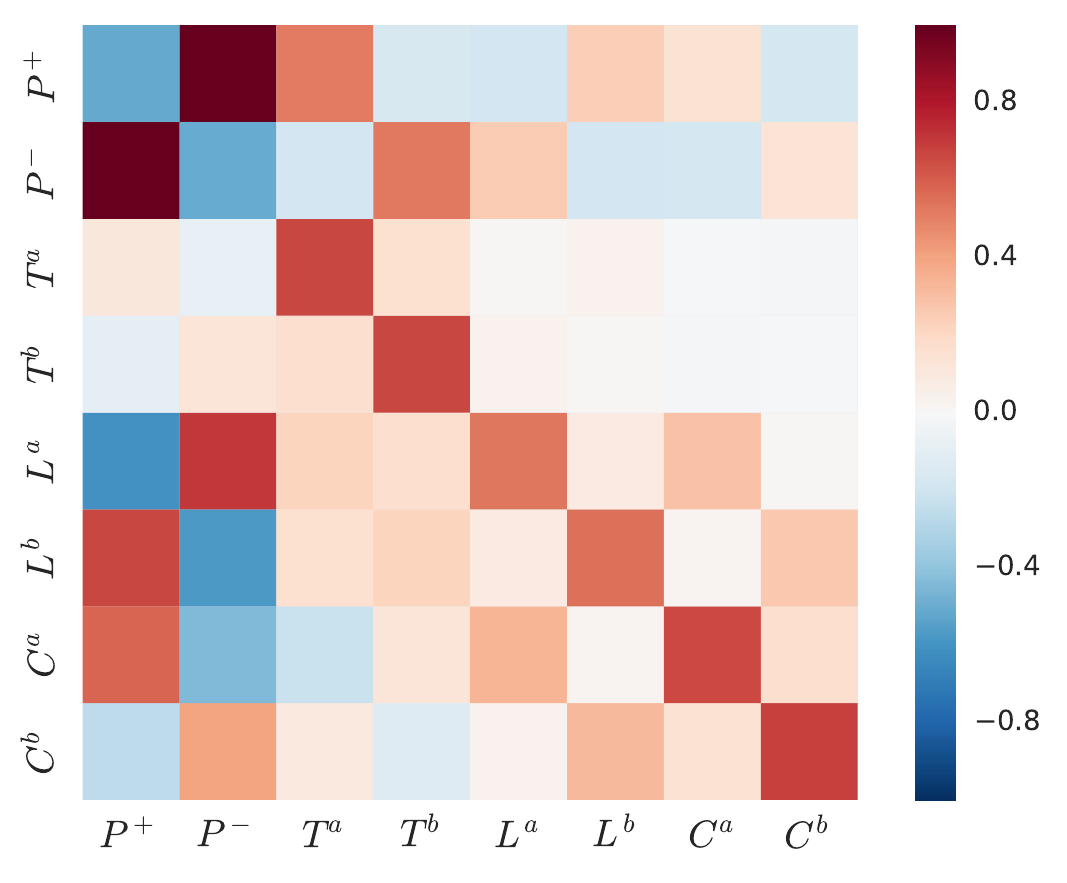}
\caption{Kernel norm matrix $\bG$ estimated with the NPHC method for the DAX future (left) and with the Wiener-Hopf method of \cite{bacry14} (right) when the 8-dimensional model described in Section \ref{sec:compwh} is considered. }
\label{fig:dax8D}
\end{figure}

In Figure \ref{fig:dax8D}, we compare the kernel integral matrices $\bG$ obtained with the NPHC method (left) with those obtained with the Wiener-Hopf approach (right) on the DAX future. Although the precise values of the matrix entries differ somewhat, as it is difficult to tune the estimation parameters of the two methods as to produce the exact same numerical results, we note that the two methods produce very consistent results. Indeed, they recover the same interaction structure and thus lead to the same interpretation of the underlying system dynamics. In our view, this represents a good sanity check for the proposed NPHC methodology. Analogous results are obtained for the Bund future. Let us also point out that the small asymmetries between symmetric interactions (such as e.g. $T^+\to T^-$ and $T^-\to T^+$) can be used get a rough measure of the estimation error. In the case presented here, the average absolute difference between symmetric interactions kernels is 0.03, which means relative error of a few percent on the most relevant interactions.

We do not comment here the features emerging from the kernel norm matrices presented in this section since they have been already discussed at length in \citep{thibault} and some of them will be further discussed in the next sections.  
Instead, here we highlight that the results of this section provide a strong case for the use of the NPHC method over the Wiener-Hopf method when the focus is solely on the kernel interaction matrix. Indeed, in order to estimate the kernel norm matrix with the Wiener-Hopf method, the full kernel functions have to be estimated first and then numerically integrated. The NPHC method thus represents a much faster alternative, as it does not require the estimation of $d^2$ functions but directly estimates their integrals. Besides the speed gain,
the gain in complexity allows NPHC to scale much better when increasing the dimension, i.e., when using more detailed models.

%
%


\subsection{A 12-dimensional mono-asset model}

By estimating directly the norm of the kernels and not the whole kernel function, the NPHC method can be used to investigate
systems of greater dimension. In this section we extend the model of Section \ref{sec:compwh} to 12 dimensions by separating the type of events that lead to a price move. The 12 even types we consider are thus $T^+$ ($T^-$), $L^+$ ($L^-$), $C^+$ ($C^-$), $T^a$ ($T^b$), $L^a$ ($L^b$), $C^a$ ($C^b$). We then apply the NPHC algorithm to estimate the branching ratio matrix. When not otherwise specified, we set $H=500s$. To further assess the validity of our methodology and the impact of time-of-day effects, we first estimate the model using different time slots within the trading day. In Section \ref{sec:12D} we also check the robustness of our results
as respect to the choice of the parameter $H$.


\subsubsection{Kernel stability during the trading day}\label{sec:facts}

We ran our method for the DAX future on the 12-dimensional point process detailed above on different subintervals of the trading day.
More precisely, we divided each trading day into 7 slots with edges at 08:00 am, 10:00 am, 12:00 am, 02:00 pm, 04:00 pm, 06:00 pm and 10:00 pm. We then estimated the 12-dimensional model described above on each slot separately, averaging over all 338 trading days available in our dataset.

\begin{figure}
\centering
\includegraphics[width=.32\textwidth]{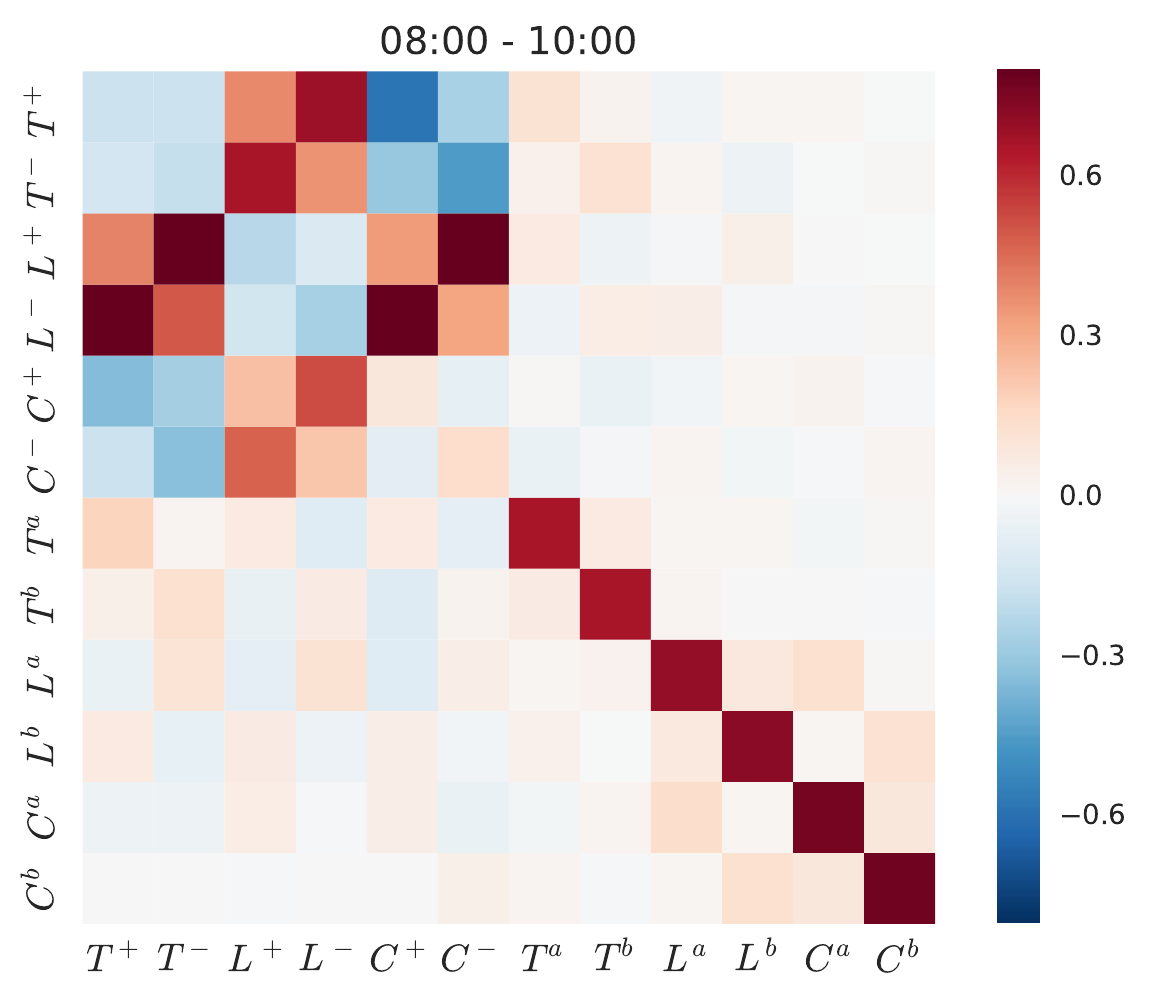}
\includegraphics[width=.32\textwidth]{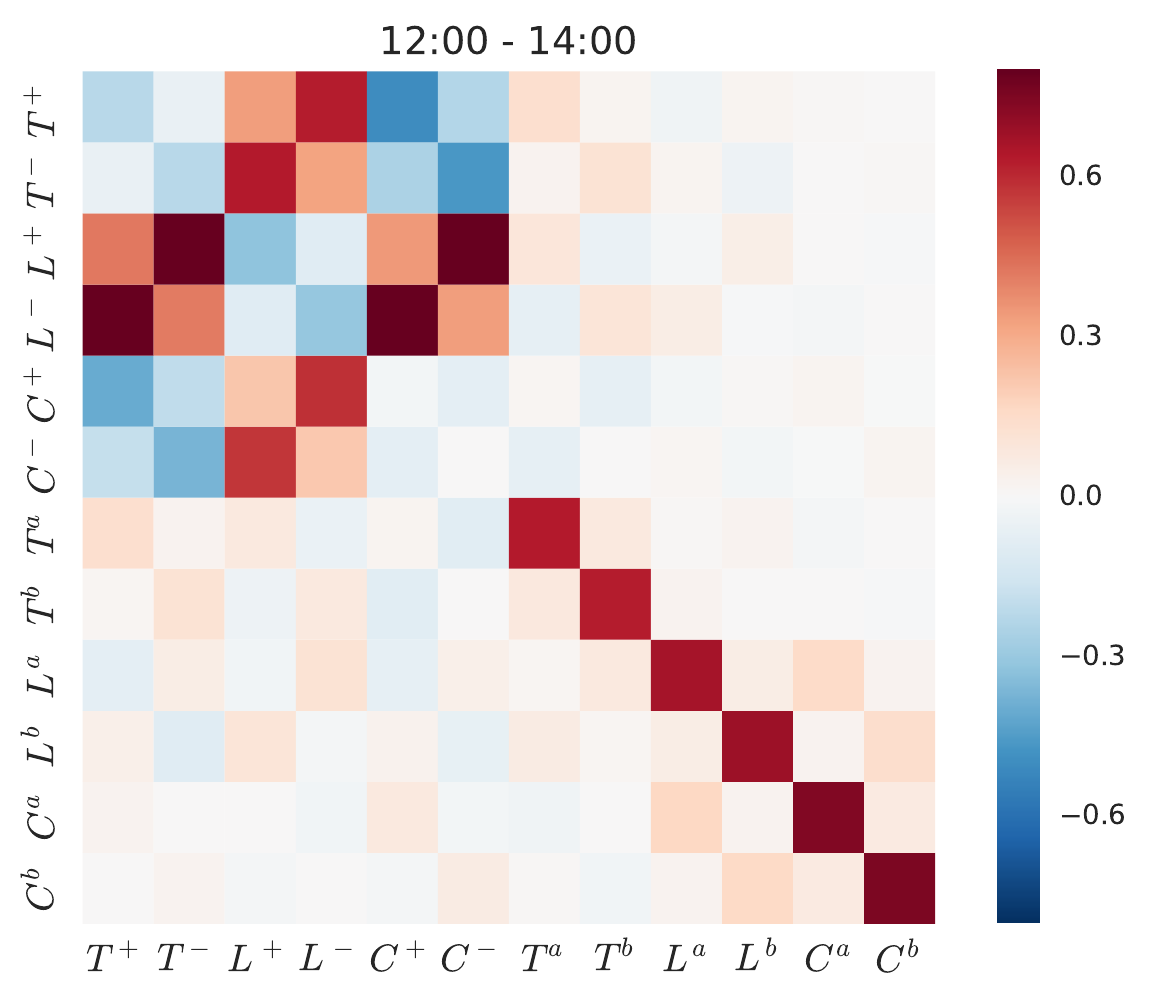}
\includegraphics[width=.32\textwidth]{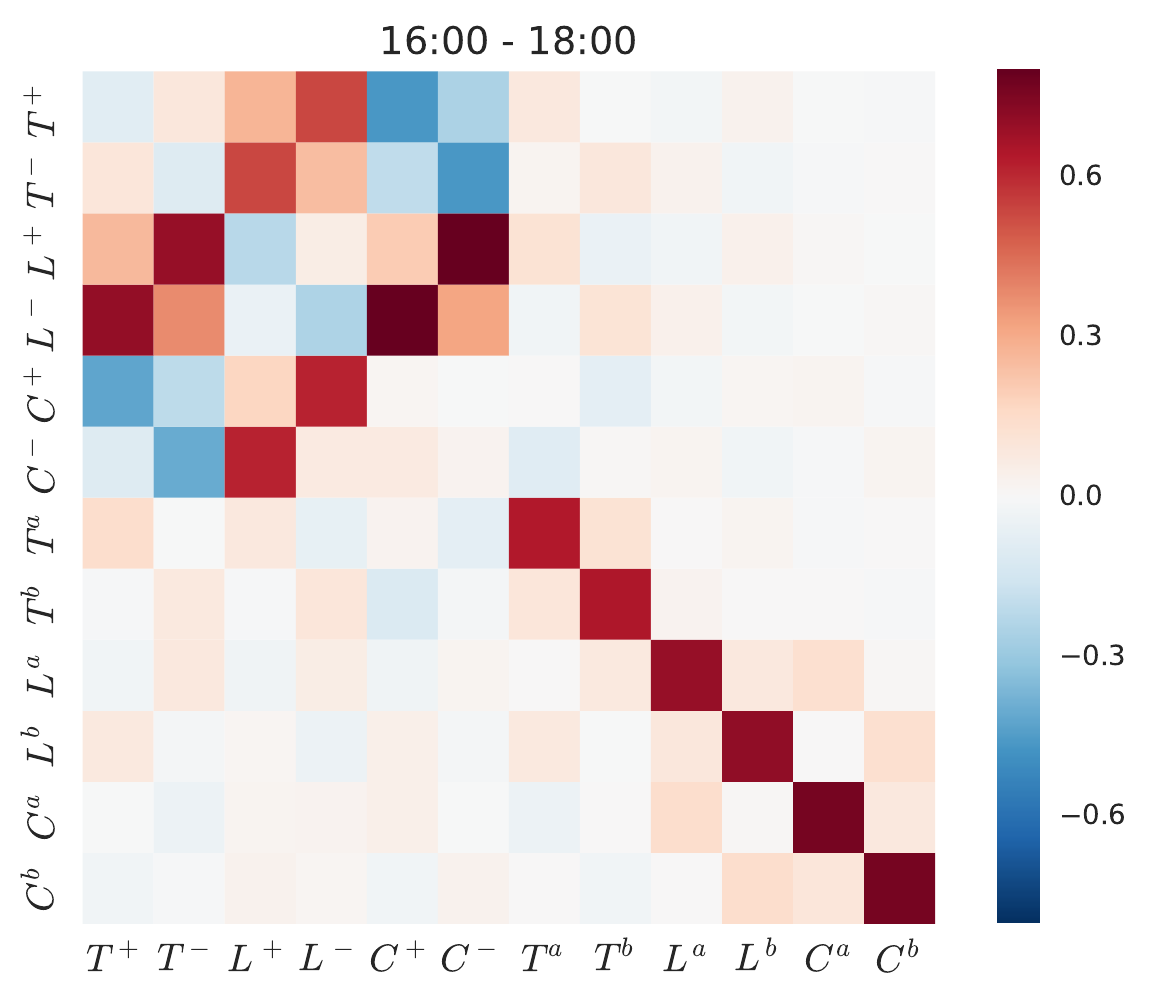}
\caption{Kernel norm matrix $\bG$ for the DAX future estimated (using NPHC) at three different times: between 08:00 and 10:00 (left), between 12:00 and 14:00 (middle) and between 16:00 and 18:00 (right). }
\label{fig:G_at_different_times}
\end{figure}

In Figure~\ref{fig:G_at_different_times} we show the estimated branching ratio matrix $\bG$ on three different slots. The results are remarkable in that the kernel norm matrix appears to be very stable during the trading day. (we checked that this is also true if we set $H=1s$).

The NPHC method outputs the estimated matrix $\widehat{\bR}$ (and then $\widehat{\bG}$)
from which one can obtain an estimate of $\bmu$ using the relation
\eqref{eq:rel_mu_lambda} that links $\bR$ and the mean intensity $\bLam$, namely $\widehat{\bmu} = \widehat{\bR}^{-1} \widehat{\bLam}$.

\begin{figure}[bp]
\centering
\includegraphics[width=0.48\textwidth]{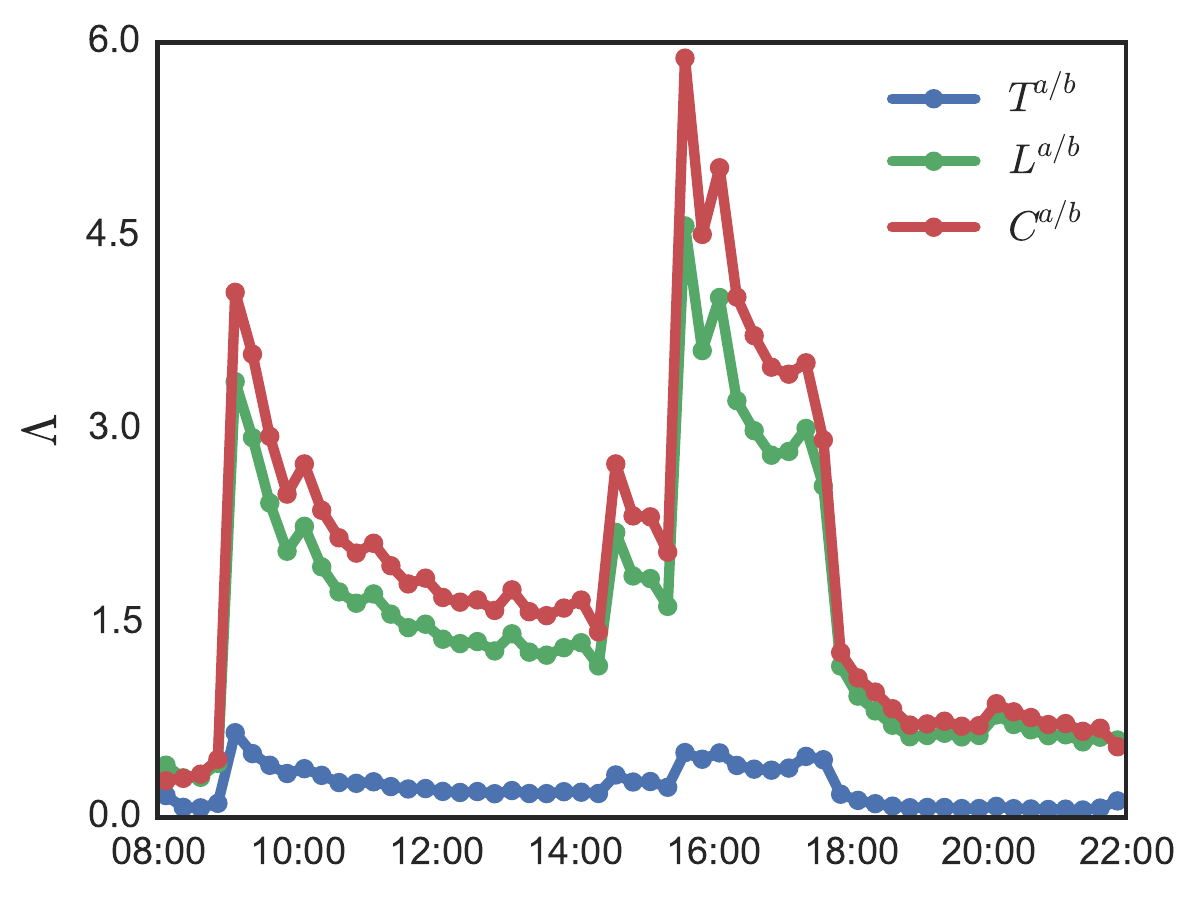}
\includegraphics[width=0.48\textwidth]{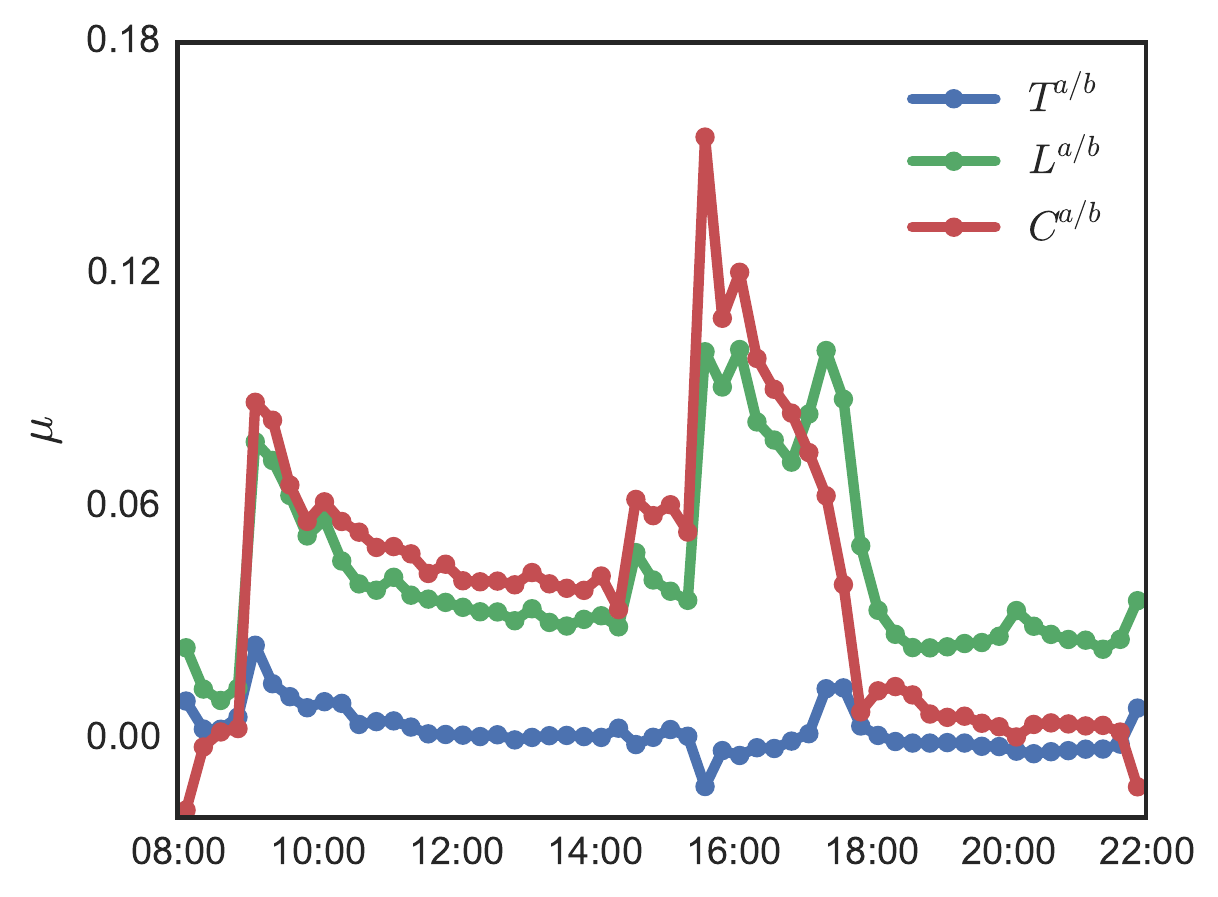}
\caption{Estimation of the baseline intensities of each event type within a trading day for the DAX future using 15 min slots. Left panel: Empirical intraday pattern measured using market, limit and cancel orders that do not move the price. Right panel: $\mu$ values estimated using the NPHC method. All quantities are expressed in $s^{-1}$.}
\label{fig:baselines}
\end{figure}

In the right panel of Figure~\ref{fig:baselines} we plot the values of $\widehat{\mu}$ as obtained using the above relation for the $T^{a/b}$, $L^{a/b}$ and $C^{a/b}$ components. We consider the kernel norm matrix as constant in each two hours slot and we estimate the average intensity on 15 minutes non-overlapping windows. Moreover, for each type of events we show the average of the bid/ask components. For comparison, in the left panel of Figure~\ref{fig:baselines} we show the empirical intraday pattern obtained for each component. We remark that the values of $\mu$ obtained with our procedure vary during the day and roughly follow the intraday curve of the respective components. Let us notice that $\mu^i/\Lambda^i$, the fraction of exogenous events, is of the order of a few percent. This is fully consistent
with what was found in \cite{thibault} and means, within the Hawkes framework, that most of the observed order book dynamics is strongly endogenous. For the price moving components the values of $\Lambda$ are of the order of 1$s^{-1}$, while results for $\mu$ are more noisy, similarly to those of $T^{a/b}$. 

This analysis confirms the result formerly observed in \cite{Bacry:2014aa} that the kernels are stable during the day, and that time-of-day effects are well captured by the baseline intensity, at least as long as we are mainly concerned with the high frequency dynamics on a very liquid asset as is the case here.

\subsubsection{Analysis of the $\bG$ matrix: Unveiling mutual interactions between book events} \label{sec:12D}

\begin{figure}
\centering
\includegraphics[width=.48\textwidth]{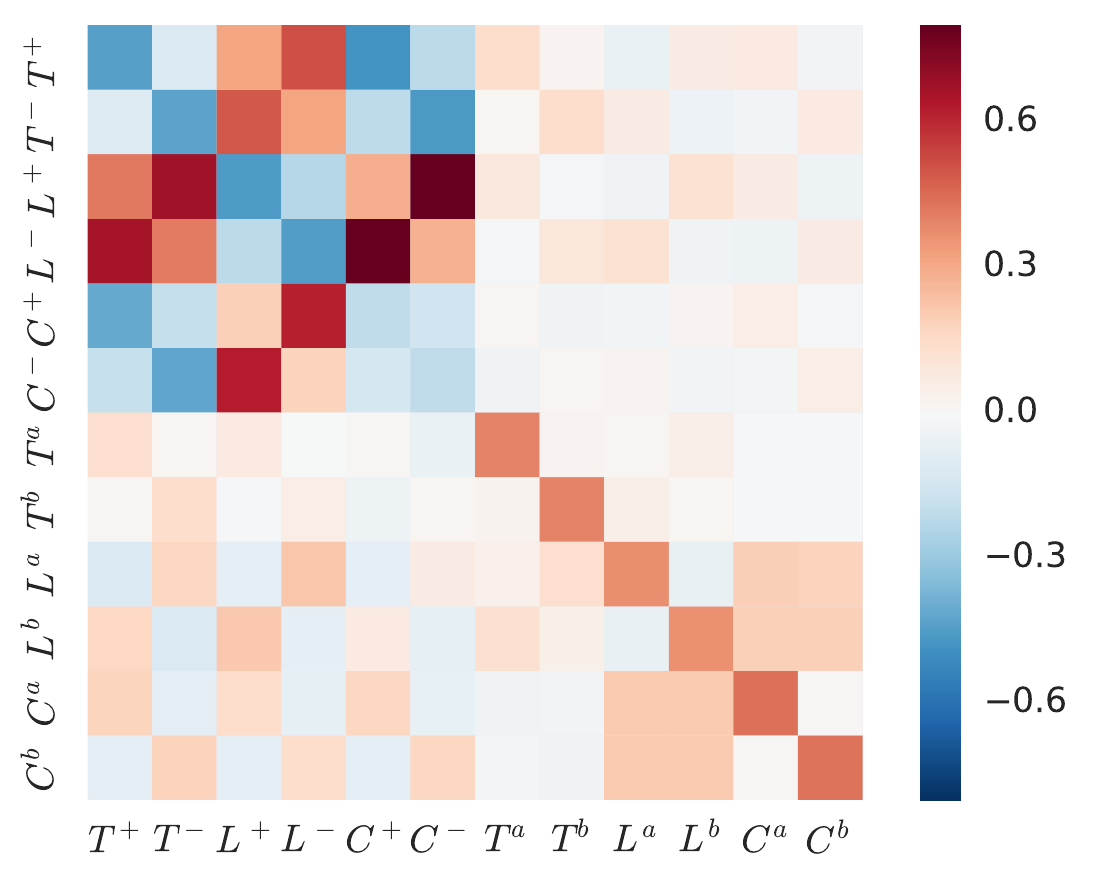}
\includegraphics[width=.48\textwidth]{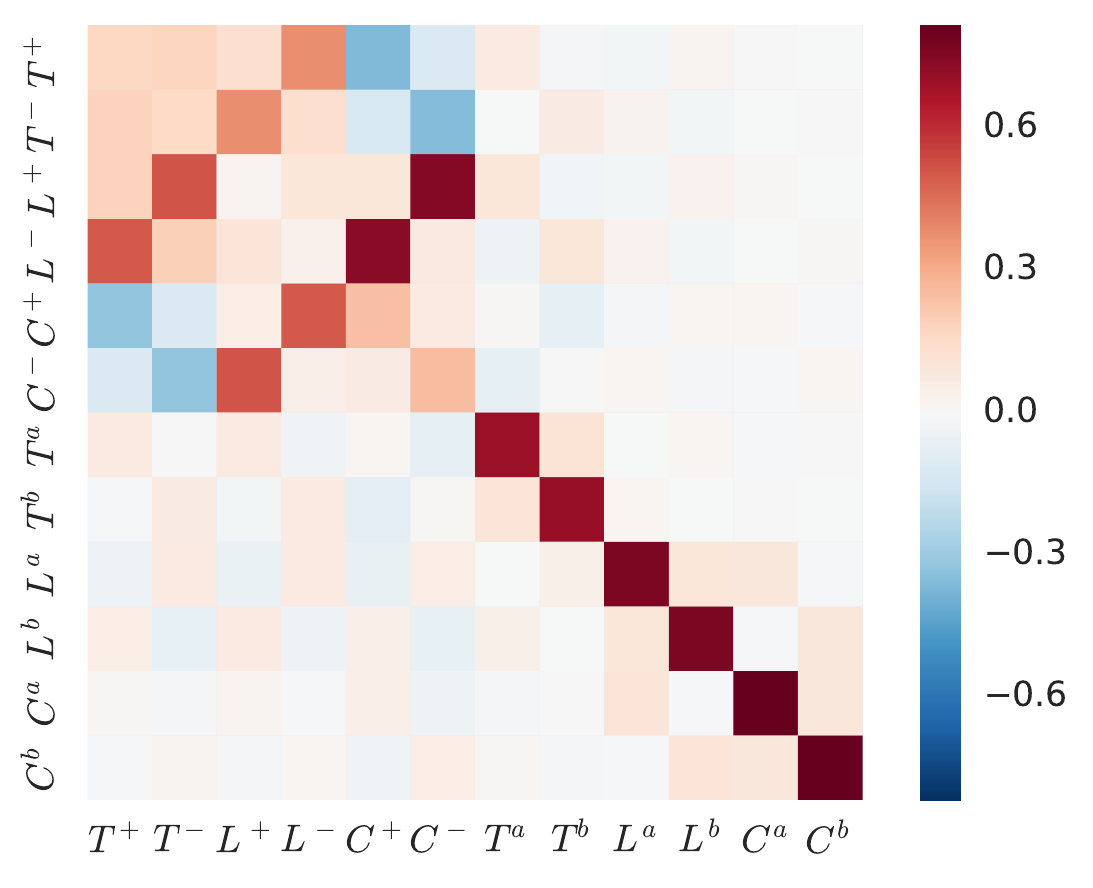}
\caption{Kernel norm matrix $\bG$ estimated with the NPHC method for the DAX future with $H=1s$ (left) and $H=500s$ (right).}
\label{fig:dax12D}
\end{figure}

Having established that the estimated kernel matrix is stable with respect to time of the day effects, we now examine more in-depth its structure. In Figure \ref{fig:dax12D} is represented the result of the estimation of the matrix $\bG$ over the whole trading day for the DAX future. The branching ratio matrix on the left panel is estimated with $H=1s$ while the right panel corresponds to $H=500s$. Let us recall that both horizons are several orders of magnitude larger than the typical inter-event time.

Concerning the differences between the two matrices, we note that certain inhibitory effects that are visible for $H=1s$ are less intense or disappear when $H=500s$ is used. This most notably happens for the elements $T^+ \to T^+$ and $T^+	\to T^-$ and similarly for $L^+\to L^{+/-}$ and $C^+\to C^{+/-}$, which suggests that when we look at longer scale correlation the self-exciting behavior (i.e. trades are followed by more trades) tends to prevail on the high frequency mean reverting effect.

Apart from these differences, we can make some observations that are valid in both cases. In particular, we note that two main interaction blocks stand out. The first is the upper left corner which concerns interactions between price-moving events, where two anti-diagonal bands are prominent. The second is the bottom right corner, which has a strong diagonal structure. The blocks  involving interactions between price-moving and non-price moving events present instead much smaller values. In what follows, we first discuss more in depth the effects of price movements on other events, then those of non-price-moving ones. We also remark that the spectral norm of the estimated matrices $\bG$ is close to $1$ while being inferior (e.g. $0.98$ for the DAX with $H=500s$). This is in line with what was found in \cite{thibault}, and the criticality of financial markets highlighted in \cite{hardiman2013critical}.

Before entering into more details, let us remark that in both cases the expected symmetry up/down ($+/-$) and bid-ask ($b/a$) is well recovered in our results. Therefore, to make notation lighter and facilitate the exposition, we will comment only on one side. More precisely, when discussing the effects of price moves we will refer only to the upwards ones ($T^+/L^+/C^+$) and when discussing effects of liquidity changes we will focus on ask side events ($T^a/L^a/C^a$).


\paragraph{Effect of price-moving events}
As we noted above, the most relevant interactions involving $T^+$ are the $T^{+} \to L^{+}$ and $T^{+} \to L^{-}$ ones, the mean reverting one ($T^{+} \to L^{-}$) being more intense. When a market order consumes the liquidity available at the best ask, two main scenarios can occur for the mid price to change again, either the consumed liquidity is replaced, reverting back the price (mean-reverting scenario, highly probable) or the price moves up again and a new best bid is created.

Market orders that move the mid price have also an inhibitory effects at short time scales on subsequent price-moving trades ($T^+ \to T^+$ is negative for $H=1s$). Indeed, once a market order consumes the liquidity available at the best quote, it is unlikely that the price will be moved in the same direction by other market orders as the price becomes more unfavorable. We also note a generally inhibitory effect of $T^+$ on price-moving cancel orders which can be linked to a mechanical effect, liquidity that has been consumed by the market order cannot be canceled anymore.

The same kind of dynamics is at play also in the interactions $L^+\to T^+$ and $L^+\to T^-$ with the roles inverted. Again, the mean reverting effect $L^+\to T^-$ appears to be much more probable. A strong mean-reverting effect is found in the block $L^+ \to C^-$. This is possibly the signature of high-frequency strategies whereby agents place limit orders in the spread and cancel them shortly thereafter.

Concerning $C^+$ events, the main feature lies in the block $C^{+} \to L^{-}$, where we notice the same anti-diagonal dominance found for the block $L^{+} \to C^{-}$. Again, we can suppose that when a limit order in the spread is removed it is often quickly replaced by market participants.

Finally, the effect of price moving events on non-price moving ones can be summarized in two main effects. The first is a trend-following/order splitting effect by which e.g. trades at the ask are likely to be followed by more trades in the same direction ($T^+\to T^a$) and similarly for limit ($L^+\to L^b$) and cancel ($C^+\to C^a$) orders. The second is the shift in liquidity triggered by a price change. A trade at the ask that moves upward the mid price triggers limit orders on the opposite side ($T^+\to L^b$). This can be understood using a latent price argument (\cite{rosenbaum_latent}),  as it is well known that there are more limit orders far from the latent price. Right after the mid price goes up, the latent price is expected to be closer to the newly best ask price than to the best bid price, thus limit order flow is expected to be higher at best bid than at best ask.

%

\paragraph{Effect of non-price-moving events}
For all events $T^a$, $L^a$ and $C^a$ the most visible feature is the strong self-exciting interaction. This has been confirmed in several works  (\cite{thibault}, \cite{rambaldi2016role}) and can be traced to order-splitting strategies and herding behaviors.
Signatures of typical trading patterns can be seen also in the kernels $L^a\to C^a$, $L^a\to C^b$, where the positive value of the kernel arises form agents canceling and replacing their limit orders with or without switching sides.

We also note the positive effects $T^a\to T^+$, $L^a\to T^-$ and $C^a\to T^+$. All these effects, as well as the analogous ones on $C^{+/-}/L^{+/-}$, reflect the fact that changes in the imbalance have an influence on the probability of a subsequent price move.
So when the queue at the best ask decrease an upward price move becomes more likely and vice-versa. These effects are much more relevant on a small tick asset (DAX) than on a large tick asset (Bund) where, the size of the queues being larger, their influence is marginal.


%

\begin{figure}
\centering
\includegraphics[width=.48\textwidth]{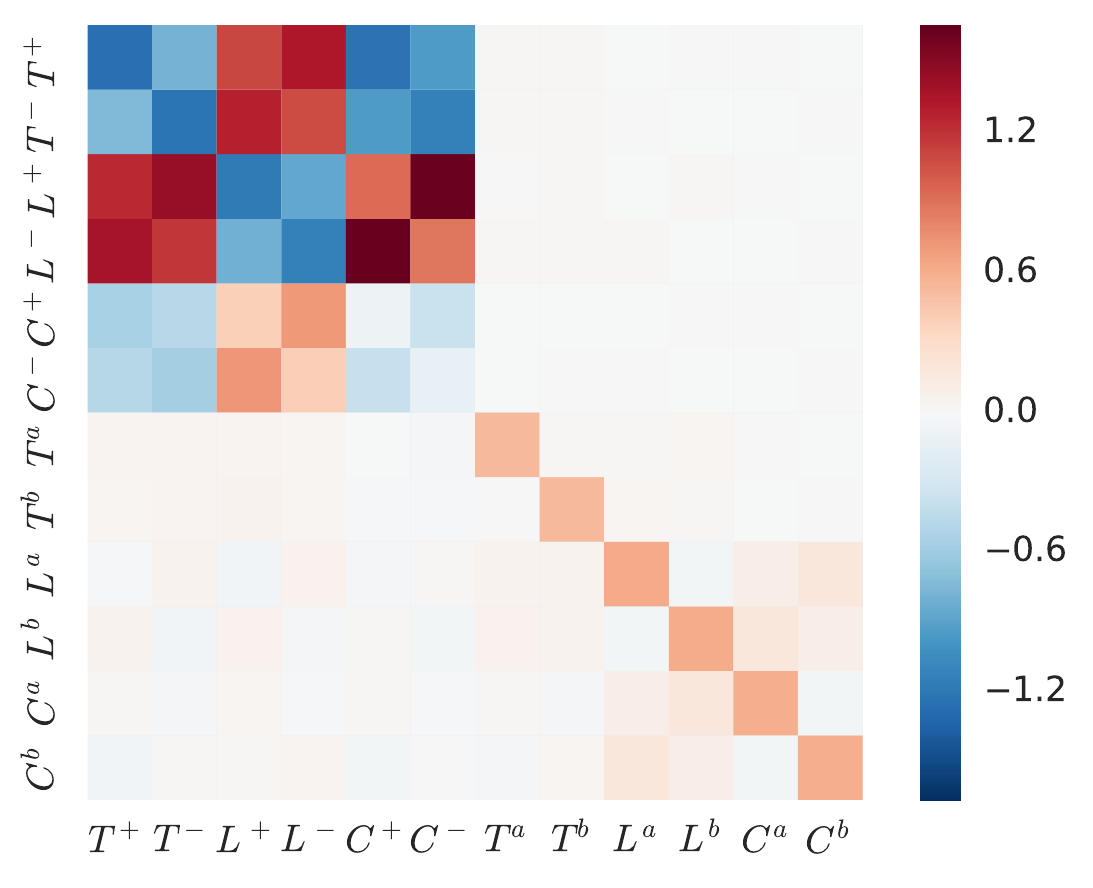}
\includegraphics[width=.48\textwidth]{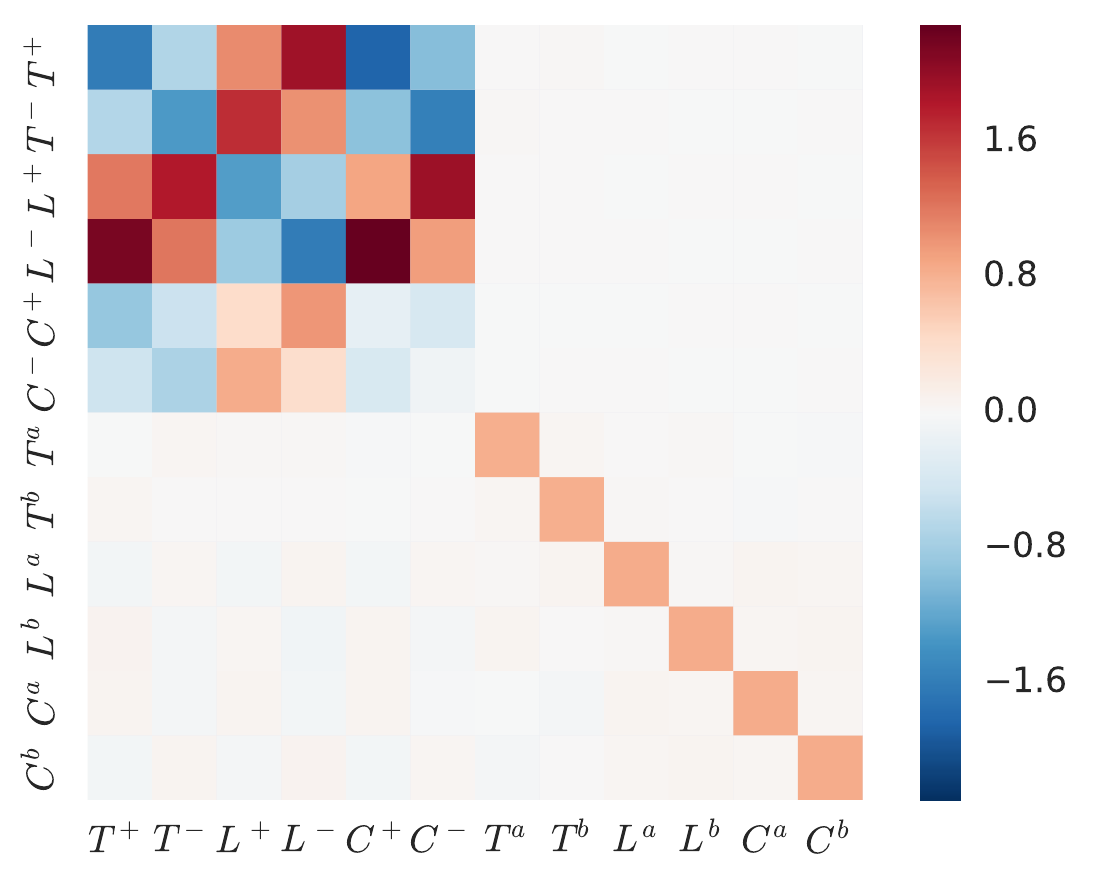}
\caption{Kernel norm matrix $\bG$ estimated with the NPHC method for the Bund future with $H=1s$ (left) and $H=500s$ (right).}
\label{fig:bund12D}
\end{figure}

We performed the same analysis on the Bund (see Figure~\ref{fig:bund12D}). The main differences as compared to the DAX are that the effects between events that move the price are much more intense while the effects of events that do not move the price on those that do move the price (and vice-versa) are much less pronounced, indeed they are barely visible in Figure~\ref{fig:bund12D}. This can be basically seen as a simple consequence of the Bund future being large tick assets, while the DAX is a small tick one. Therefore, price movements on the former are much less frequent but when they happen their effects are more marked.

\subsubsection{Analysis of the $\boldsymbol{\Psi}$ matrix: the fingerprint of meta-orders}
\begin{figure}
\centering
\includegraphics[width=.48\textwidth]{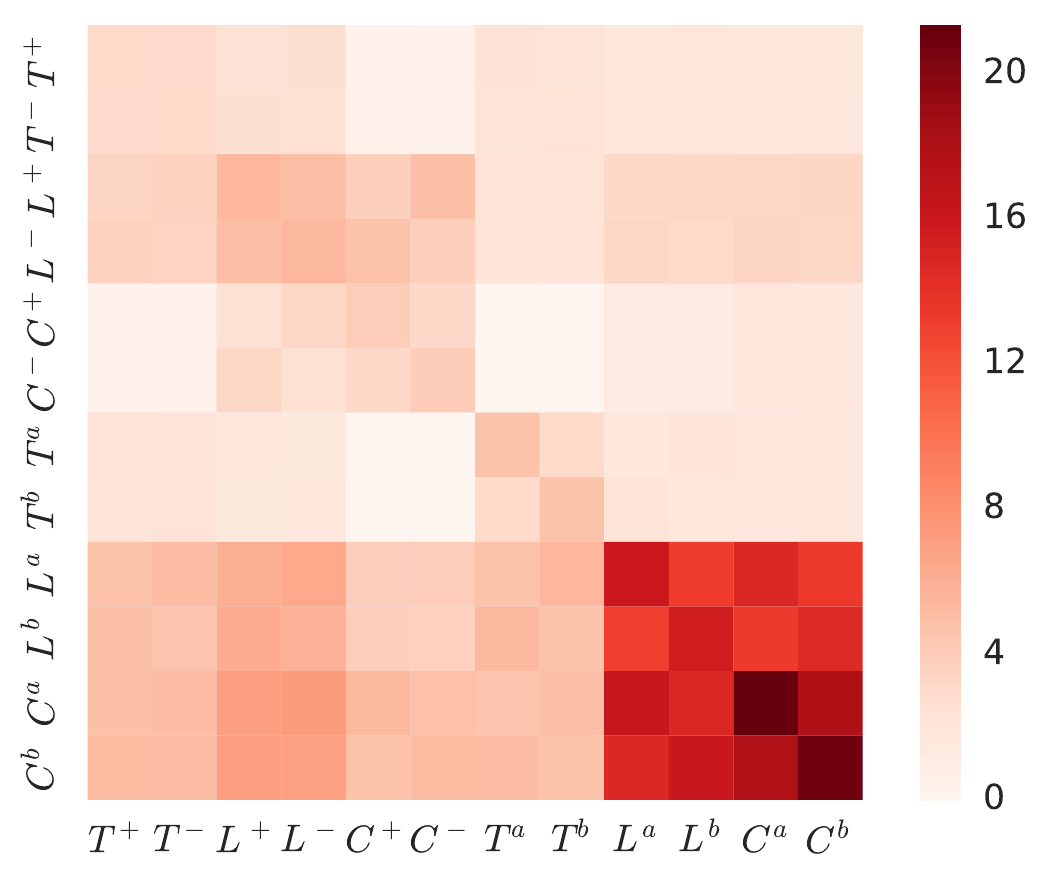}
\includegraphics[width=.48\textwidth]{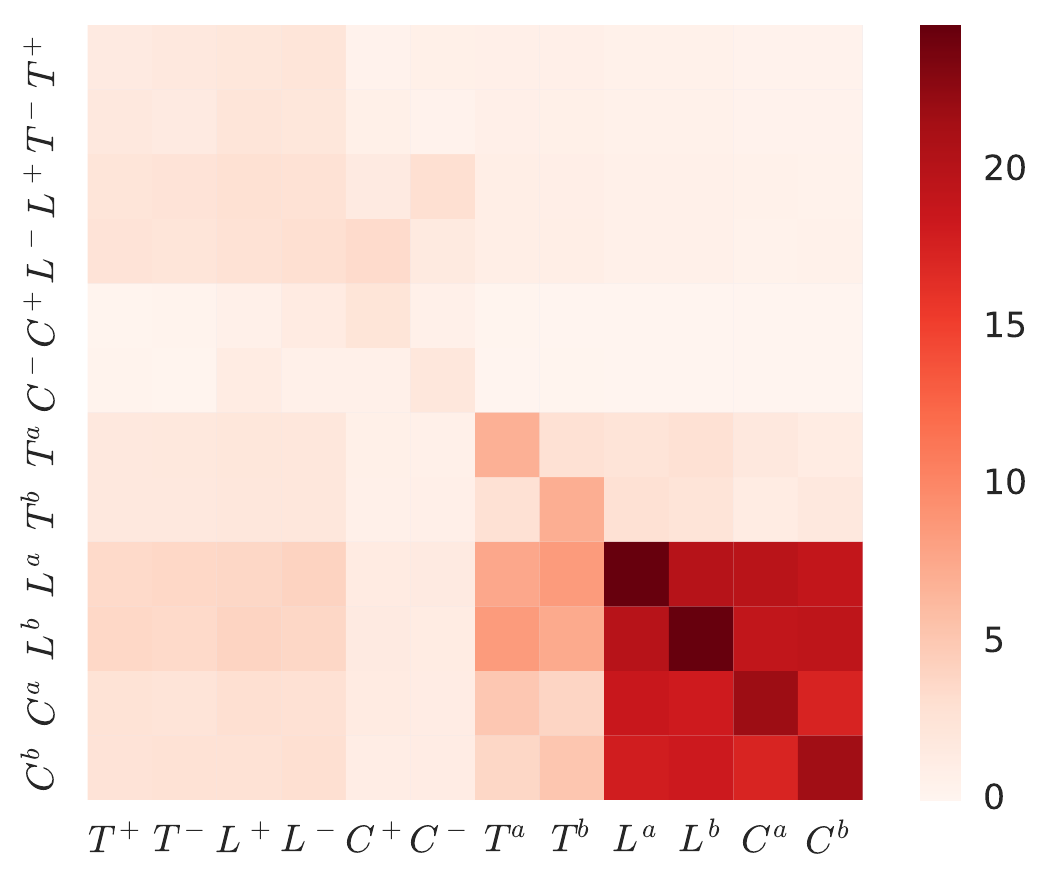}
\caption{$\boldsymbol{\Psi}$ matrix of eq. \eqref{eq:defpsi} estimated with the NPHC method for the DAX future (left) and the Bund future (right) with $H=500s$.}
\label{fig:psi}
\end{figure}

As discussed in Section~\ref{sec:Hawkes}, the elements of the matrix $\boldsymbol{\Psi}$ quantifies the total effect, direct and indirect, of an event of type $j$ on events of type $i$. More precisely, thanks to the branching process structure, we can interpret $\psi^{ij}$ as the mean number of events of type $i$ generated by a single exogenous ancestor of type $j$.
We plot the estimated matrices $\boldsymbol{\Psi}$ for the DAX and Bund futures in Figure~\ref{fig:psi}. The main feature that appears for both assets is the set of strong values found in the bottom right corner, namely in the columns and lines associated with $L^{a/b}$ and $C^{a/b}$.
We note that an exogenous limit or cancel event generates a large number of limit and cancel events and,
to a lesser extent, trade events.
This can be read as the signature of meta-orders. Indeed, if an agent wants to sell a large number of contracts{\footnote{Let us recall that, in our discussion, we only address half of the matrix coefficients since the discussion on the other half can be obtained using the symmetries ask/bid, buy/sell, price up/price down. Following these lines, we only consider here the case of a selling meta-order.}}, he will place a meta-order, i.e., he will optimize the overall cost by dividing this large order into several smaller orders. The overall optimization will result in many limit/cancel sell orders $L^a,C^a$ and, as less as possible, of sell market orders $T^b$ (the cost of a market order is on average higher than that of a limit order). The same description can be applied to understand why
an exogenous sell market order $T^b$ generates mainly limit and cancel sell orders $L^a,C^a$ as well as other sell market orders $T^b$.


Due to the much lower values of the exogenous intensities for price moving events, the left part of the $\boldsymbol{\Psi}$ matrix is more noisy. Nevertheless, at least in the DAX case, we note also for the price moving components the prevalence of the $L^+\to L^+$ and $L^+ \to C^-$ elements, which are the price-moving counterparts of the effect described for $L^a$.

Finally, we also remark that although we noted several inhibition effects in the matrices $\bG$, the elements of $\boldsymbol{\Psi}$ are non negative. This suggests that most inhibition effects are short lived and the effect of an event arrival is towards an increase of the overall intensity. This is in line with what was found in \cite{thibault} and \cite{rambaldi2016role}, where the inhibitions effects were shown to be mostly concentrated around the typical market reaction time.

Within the branching ratio representation of Hawkes processes, $\frac{\mu^j}{\Lambda^i} \psi^{ij}$ represents the fraction of events of type $i$ that has a type $j$ as primary ancestor.
Along the same line, we can estimate the fraction of aggressive orders (i.e. all $T$),  as opposed to passive orders ($L$ or $C$), that is ultimately generated by another aggressive order, as:
\begin{equation}
\frac{1}{\sum_{ i=\lbrace T^{+/-}, T^{a/b} \rbrace} \Lambda^i} \sum_{j= \lbrace T^{+/-}, T^{a/b}\rbrace} \sum_{i=\lbrace T^{+/-}, T^{a/b}\rbrace} \psi^{ij} \mu^j \; .
\end{equation}

We find that for both assets this fraction is about 10\%, which means that the large majority of market orders have a ``passive order" ($L$ or $C$) oldest ancestor. We compute the analogous fraction for passive orders and we find that for both assets more than 96\% of the passive orders ($L$ or $C$) have an oldest ancestor that is itself a passive ($L$ or $C$) order.
This fact is in line with the idea that meta-orders would be at the origin of most of the trading activity within the order book.

\section{Multi-asset model}\label{sec:multi}
{
Studying and quantifying the interactions and comovements within a basket of assets is an important topic in finance. Most of these studies focus on the return correlations properties
in relationship with portfolio theory. At very high frequency, the discrete nature of price variations and the asynchronous occurrence of price change events make the correlation analysis trickier and, in order to avoid well known bias (like the Epps effect) one has to use specific techniques like the estimator proposed by \cite{hayashi2005covariance}. Hawkes processes, being naturally defined in continuous time, can represent a complementary tool for the investigation of high-frequency cross-asset dynamics.

The idea of capturing the joint dynamic of multiple assets via Hawkes processes has only been considered
in few recent papers. Let us mention the work proposed by \cite{bormetti2015modelling} which models the simultaneous cojumps of different assets using a one-dimensional Hawkes process, and a more recent work (\cite{da2017correlation}) which focuses on the correlation and lead-lag relationships between the price changes of two assets, in the spirit of \cite{bacry2013modelling}.

In this section, we aim at unveiling a more precise structure of the high-frequency cross-asset dynamics by pushing further the dimensionality of the model to include simultaneously events on two assets. We first consider the pair DAX-EURO STOXX and then the one Bobl-Bund. The pairs of assets considered here are tightly related, as they share exposure to the same risk factors and, in the case of DAX-EURO STOXX, also because the underlying indices actually share a significant part of their components. This is confirmed also by Table \ref{tab:corr} where we report 5 minutes return correlations among the considered assets.

In this section we consider the same kind of events as in Section \ref{sec:compwh} and we have therefore a $16$-dimensional model  ($2\times 8$) corresponding to 256 possible interactions. Let us point out that this is quite a large dimension value for a non parametric methodology.

\begin{table}[tbh]
    \centering
    \begin{tabular}{lrrrr}
\toprule
{} &   DAX &   ESXX &   Bobl &   Bund \\
\midrule
DAX &  1.00 &  0.89 & -0.18 & -0.22 \\
ESXX &  0.89 &  1.00 & -0.19 & -0.22 \\
Bobl & -0.18 & -0.19 &  1.00 &  0.85 \\
Bund & -0.22 & -0.22 &  0.85 &  1.00 \\
\bottomrule
\end{tabular}
    \caption{Five minutes return correlation coefficients for the examined assets.}
    \label{tab:corr}
\end{table}


\subsection{The DAX - EURO STOXX model}

In the following, we will denote the events of the DAX order book with the subscript $D$ while we will use the subscript $X$ for the events
of EURO STOXX order book.
The obtained branching ratio matrix is displayed in Figure \ref{fig:daxesxx_complete}. We observe that the mono-asset submatrices (the two $8 \times 8$ block matrices along the diagonal), which present the most relevant effects, have the same structure as the ones which have already been commented on in detail in Section \ref{sec:compwh}.
Consequently, in this section,  we shall focus our discussion on the non diagonal $8 \times 8$ submatrices that correspond to the interactions between the two assets. These two submatrices are shown in Figure \ref{fig:daxesx16D}. Note that colors have been rescaled to highlight their structure. To keep the notation lighter, we will comment only on effect of upwards price moves and ask events as it was done in the previous section, since we find the symmetries $+/-$ and $a/b$ to be well respected.
The most striking feature emerging from Figure \ref{fig:daxesx16D} is the very intense relation between same-sign price movements on the two assets. Albeit present in both directions, the norms $P^+_X \to P^+_D$ attain larger values.

Another notable aspect is the different effects of price moves and liquidity changes of one asset on events on the other asset. Price moves on the DAX have also an effect on the flow of limit orders on EURO STOXX ($P^+_D\to L^b_X$ and  $P^+_D\to C^a_X$), whereas EURO STOXX price moves triggers mainly DAX price moves in the same direction ($P^+_X\to P^+_D$).
An important aspect for understanding this result is the different perceived tick sizes on the two assets.

In the following, whenever it is convenient, we shall place the discussion withing the framework of latent price models
(e.g., \cite{rosenbaum_latent}). Within this framework, the latent price refers to an underlying efficient price representing at any time some average opinion of market participants about the value of the asset.
As noted in Section \ref{sec:data}, the DAX future is a  small-tick asset, while the EURO STOXX future is a large-tick one (\cite{eisler2012}).
As a consequence, an upward move in the DAX price ($P^+_D$), while signaling that the market latent price has moved slightly upwards, is not sufficient to move the EURO STOXX price by a full tick. However, this move can be perceived in the EURO STOXX through the $L_X^b$ and $C_X^a$ flows that are increasing. Indeed, as already mentioned in Section \ref{sec:compwh}, it is well known (\cite{rosenbaum_latent}) that there are more limit orders far from the latent price. The latent price went up, so it is now closer to the best ask, and hence the flow of the limit (resp. cancel) orders on the best bid (resp. ask) is increasing.



In the opposite direction, a change in EURO STOXX price is perceived as ``large" and triggers price changes in the same direction on the DAX. Interestingly, we can also note that changes in the latent price on the EURO STOXX triggers price movements on the DAX. For instance, a shift of liquidity at the bid, namely an increase of the arrival flow of limit orders at the bid, that signals that the latent price has moved upwards, has a direct effect on upward price moves on the DAX. This can be seen from the interactions $T_X^a \to P^+_D$, $L_X^b \to P^+_D$ and $C_X^a \to P^+_D$.


We can summarize our results by saying that price changes and liquidity changes on the DAX mainly influence liquidity (latent price) on the EURO STOXX, while price changes and liquidity changes on the EURO STOXX tend to trigger price moves on the DAX.

Finally, let us note that the above effects are even more pronounced when we estimate the interaction matrices with a smaller $H$. In particular the effects of DAX price movements on $T, L, C$ on the EURO STOXX become more relevant compared with those on prices. At the same time, while the effect of EURO STOXX price moves on DAX's ones is still strong, the effect of liquidity movements on DAX price movements is comparatively stronger with smaller $H$. This suggests that these effects are mainly localized at short time scales, while the $P^+\to P^+$ ones have much slower decay in time.

\begin{figure}[p]
\centering
\includegraphics[width=.75\textwidth]{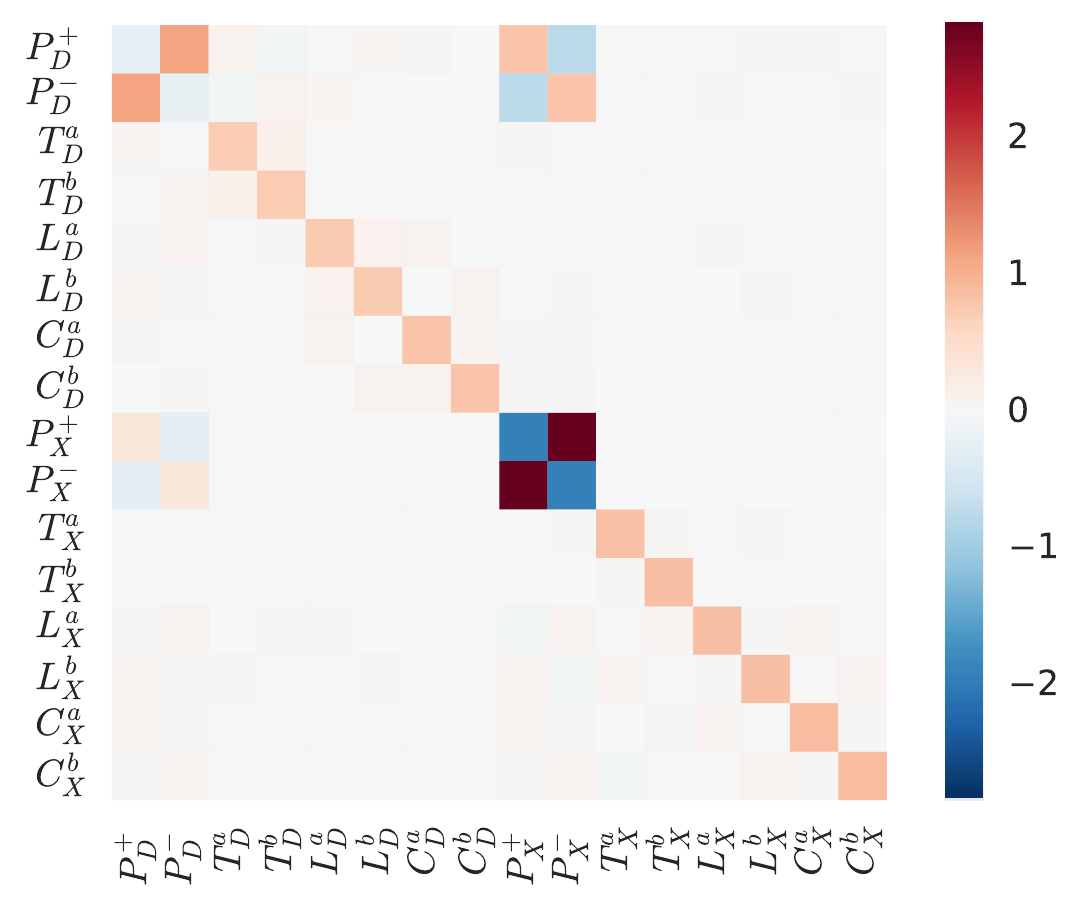}
\caption{Hawkes kernel norm matrix obtained when the DAX and EURO STOXX futures are considered simultaneously in a 16D model. DAX events are denoted with the $D$ subscript, EURO STOXX ones with the $X$ subscript.}
\label{fig:daxesxx_complete}
\end{figure}

\begin{figure}[p]
\centering
\includegraphics[width=.48\textwidth]{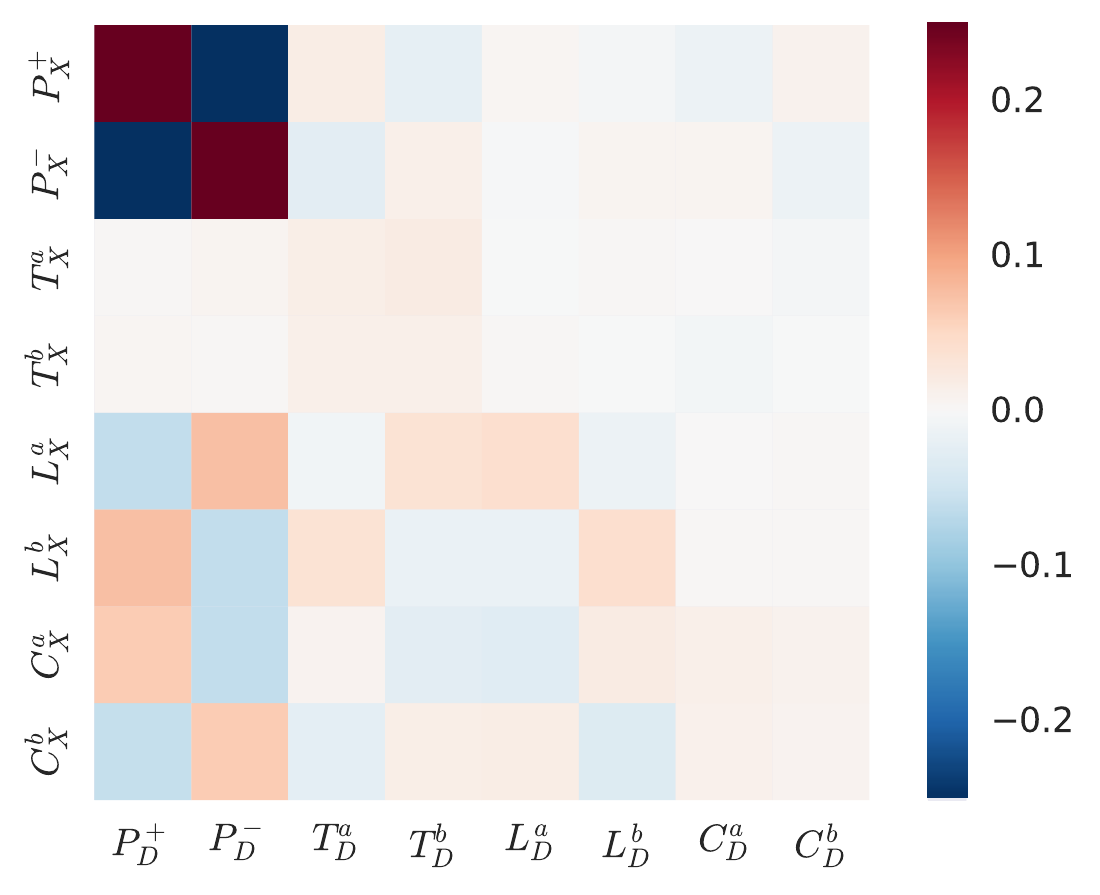}
\includegraphics[width=.48\textwidth]{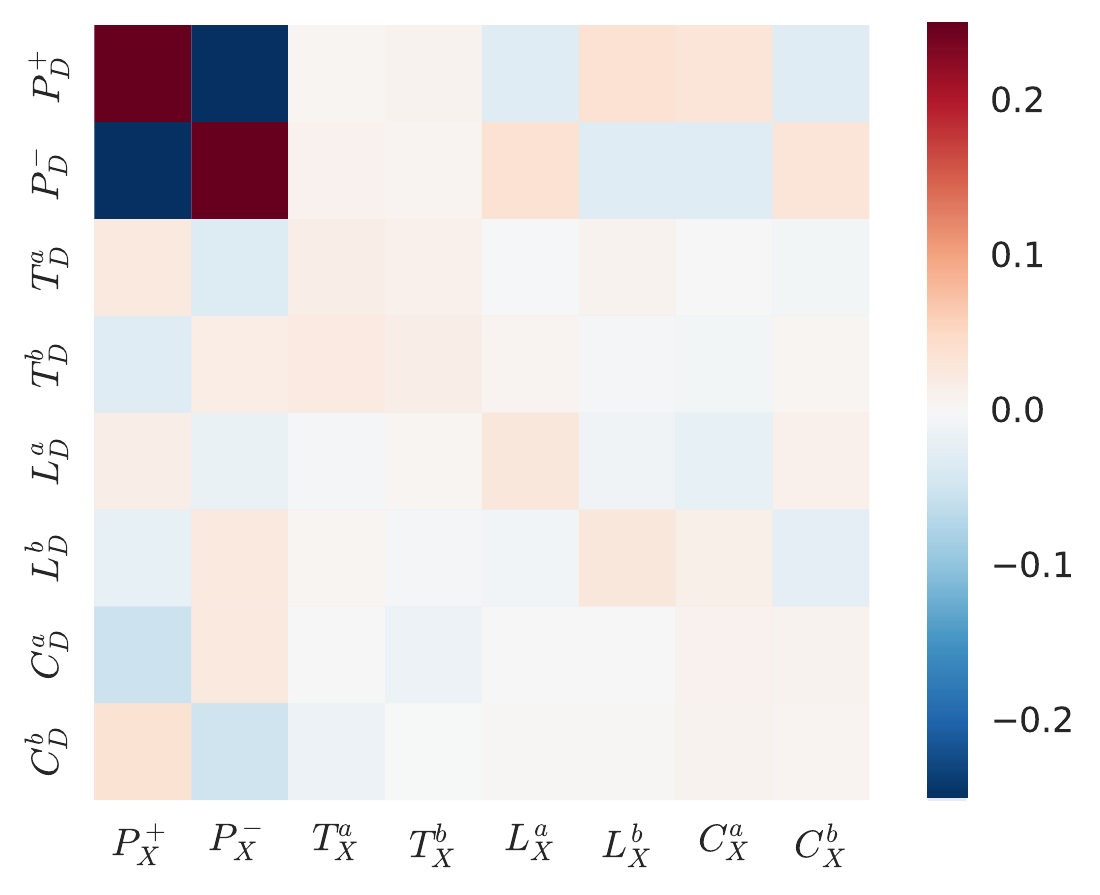}
\caption{Submatrices of the Kernel norm matrix $\bG$ corresponding to the effect of DAX events on EUROSTOXX STOXX events (left) and vice versa (right). These two submatrices correspond to the ones lying on the antidiagonal on the Figure \ref{fig:daxesxx_complete}}
\label{fig:daxesx16D}
\end{figure}

\subsection{Bobl - Bund}

We perform the same analysis on the asset pair Bobl-Bund futures. Here both assets are large tick assets, however the Bund is much more actively traded than the Bobl in the sense that all the order flows are of higher intensity.
The cross-asset submatrices are depicted in Figure \ref{fig:boblbund16D}. As in the previous case, we remark that the elements $P^+_L \to P^+_M$ and $P^+_M \to P^+_L$  reflect the strong correlation observed between the two assets. Price changes in the Bund have also a noticeable effect on limit/cancel order flows in the Bobl, while price changes in the Bobl have little to no effect on the Bund except for the mentioned $P^+_M \to P^+_L$  interaction. At the same time, $T^a, L^a, C^a$ events on the Bobl impact prices on the Bund, while the corresponding event on the Bund have little effect.

Comparing this with the case of the DAX-EURO STOXX pair, we can liken the effect of the Bund on the Bobl to that of the DAX over the EURO STOXX and vice-versa. We argue that the difference in trading frequency between the Bobl and Bund contracts has a similar effect of that of a different tick size that we observed in the previous case. As before, we have an asset, the Bund, which is more ``reactive'' (the limit/cancel order flows are higher than those of the Bobl) than the Bobl, thus  a price change of the Bund indicating a change of the latent price impacts the limit/cancel flows of the Bobl. In the previous case, the higher ``reactivity'' of the DAX was due to its smaller tick size.

\begin{figure}[p]
\centering
\includegraphics[width=.48\textwidth]{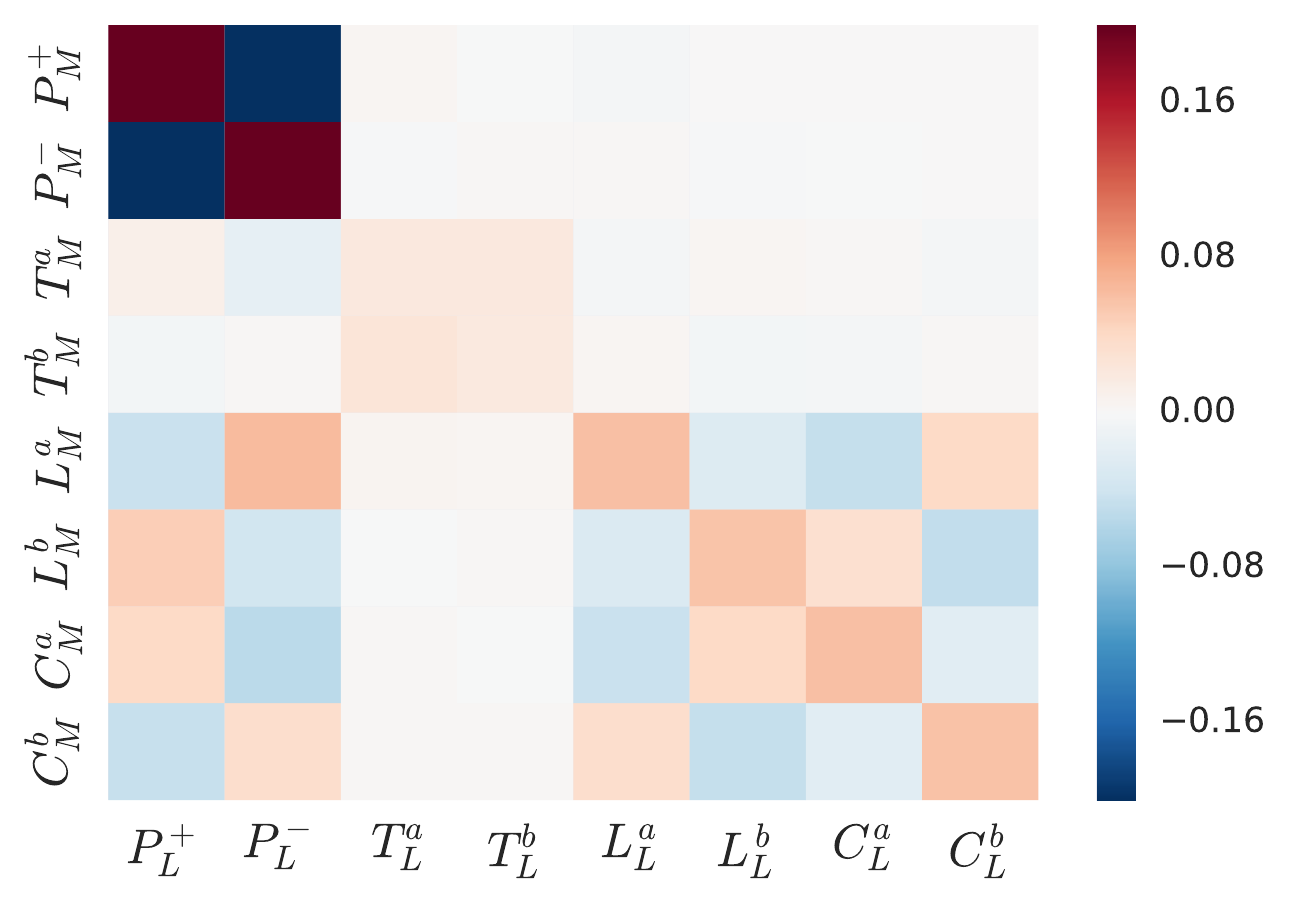}
\includegraphics[width=.48\textwidth]{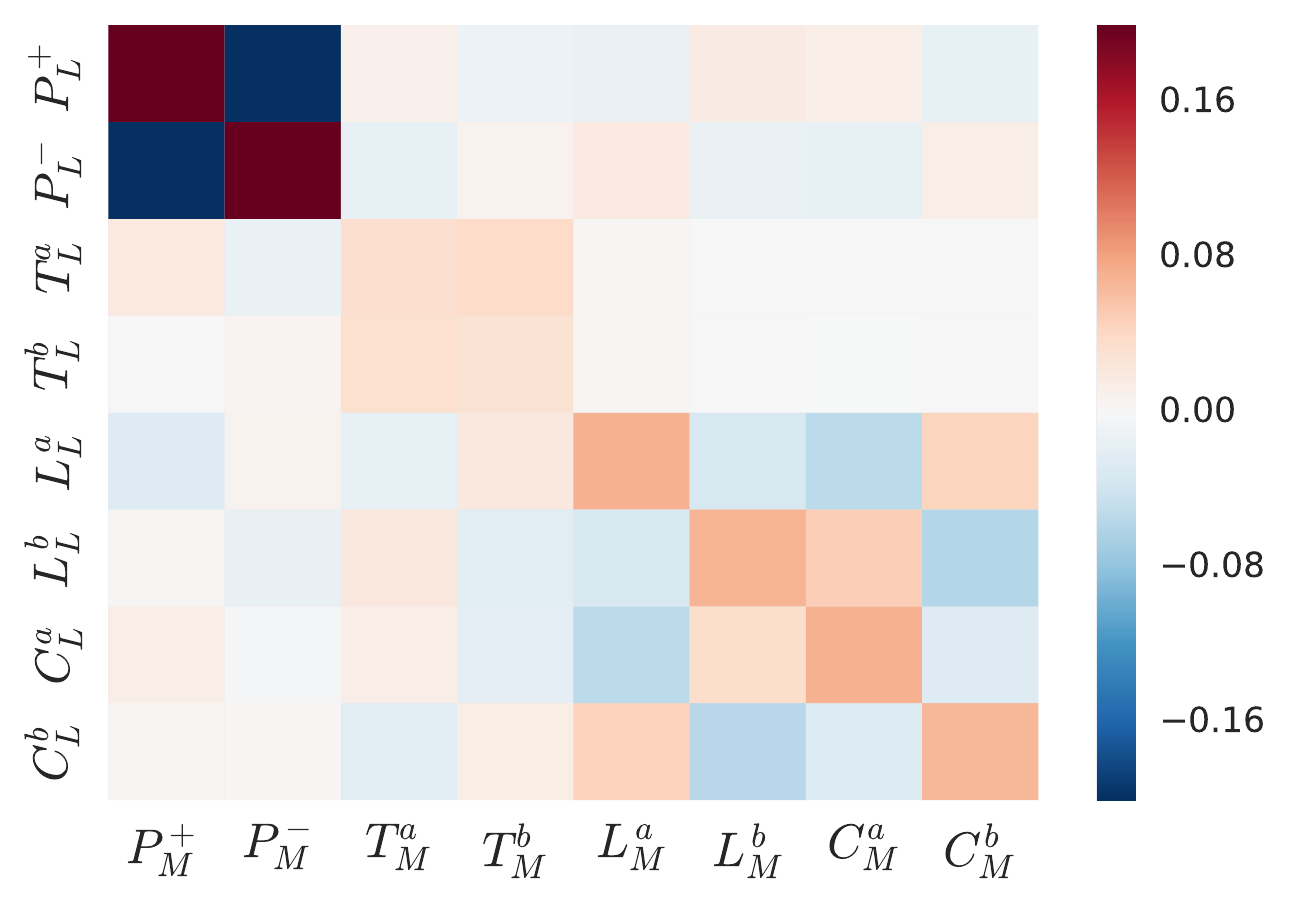}
\caption{Submatrices of the Kernel norm matrix $\bG$ corresponding to the effect of Bund ($L$) events on Bobl ($M$) events (left) and vice-versa (right).}
\label{fig:boblbund16D}
\end{figure}

\section{Conclusion and prospects}
\label{sec:conclusion}

In the context of Hawkes processes, the estimation of the matrix kernel norms is essential, as it gives a clear overview of the dependencies involved in the underlying dynamics. In the context of high-frequency financial time-series non-parametric estimation of the matrix kernel norms has already shown to be very fruitful (\cite{Bacry:2014aa,thibault}), since
it provides a very rich summary of the system interactions, and it can thus be a valuable tool in understanding a system where many different types of events are present.
However, its estimation is a computationally demanding process since these estimations are computed from a non-parametric pre-estimation of the kernels themselves, i.e., their entire shape and not only their norm. The resulting complexity prevents the estimations from being performed when the dataset is too heavy or (more important) when the dimension of the Hawkes process (i.e., the number of considered different event types) is too large.

In this work, we presented the newly developed NPHC algorithm (\cite{achab2016uncovering}) that allows to {\em directly} estimate non-parametrically the kernel norms matrix of a multidimensional Hawkes process, that is without going through the kernel shapes pre-estimation step. As of today, it is the only {\em direct} non-parametric estimation procedure available in the academic literature. This method can be seen as a Generalized Method of Moments (GMM) that relies on second-order and third-order integrated cumulants.
This paper shows that this method successfully reveals the various dynamics between the different (first level) order flows involved in order books. In a context of a single-asset 8-dimensional Hawkes process, we have shown (as a ``sanity check'') that it is able to reproduce former results obtained using ``indirect'' methods. Moreover, the so-obtained gain in complexity allowed us to run a much more detailed analysis (increasing the dimension to 12), separating the different types of events that lead to a mid-price move.
This in turn allowed us to have a very precise picture of the high frequency order book dynamics, revealing, for instance, the different interactions that lead to the high-frequency price mean reversion or those between liquidity takers and liquidity makers as well as the influence of the tick-size of these dynamics.
Not the least, through the analysis of the matrix $\boldsymbol{\Psi}$ we also detected the signature of meta-orders.
We have also successfully used the NPHC algorithm in a multi-asset 16-dimensional framework. It allowed us to unveil very precisely the high-frequency joint dynamics of two assets that share exposure to the same risk factors but that have different characteristics (e.g., different tick sizes or different degrees of reactivity).
It is noteworthy that our methodology can efficiently highlight these types of dynamics, especially since cross-asset effects are second order effects compared to mono-asset's.

We conclude by noting that our study left out some relevant information such as the volume of the orders and the size of the jumps in the mid-price. This will be the objective of future works.
Moreover, within the methodology presented in this paper, an analysis of baskets of assets (with more than two assets) as well as multi-agent high-frequency interactions are currently under progress.

\section*{Acknowledgments}
This research benefited from the support of the Chair ``Changing Markets'', under the aegis of Louis Bachelier Finance and Sustainable Growth laboratory, a joint initiative of Ecole Polytechnique, Universit\'e d'Evry Val d'Essonne and F\'ed\'eration Bancaire Fran\c{c}aise and from the chair of the Risk Foundation: Quantitative Management Initiative.

\bibliography{biblio.bib}

\bibliographystyle{rQUF}

\appendix

\section{Origin of the scaling coefficient $\kappa$}
\label{appen:origin_kappa}

Following the theory of GMM, we denote $m(X,\theta)$ a function of the data, where $X$ is distributed with respect to a distribution $\mathbb{P}_{\theta_0}$,
which satisfies the \emph{moment conditions} $g(\theta) = \mathbb{E}[m(X,\theta)] = 0$ if and only if $\theta = \theta_0$, the parameter $\theta_0$ being the \emph{ground truth}.
For $x_1, \ldots, x_N$ observed copies of $X$, we denote $\widehat{g}_i (\theta) = m(x_i, \theta)$,
the usual choice of weighting matrix is $\widehat{W}_N (\theta) = \frac{1}{N} \sum_{i=1}^N \widehat{g}_i (\theta) \widehat{g}_i (\theta)^\top$,
and the objective to minimize is then
\begin{align}
\left( \frac{1}{N} \sum_{i=1}^N \widehat{g}_i (\theta) \right) \left( \widehat{W}_N (\theta_1)\right)^{-1} \left( \frac{1}{N} \sum_{i=1}^N \widehat{g}_i (\theta) \right),
\end{align}
where $\theta_1$ is a constant vector. Instead of multiplying by the inverse weighting matrix, we have decided to divide by the sum of its eigenvalues, which is easily computable:
\begin{align*}
\mbox{Tr}(\widehat{W}_N (\theta)) &= \frac{1}{N} \sum_{i=1}^N \mbox{Tr}(\widehat{g}_i (\theta) \widehat{g}_i (\theta)^\top) \\
&= \frac{1}{N} \sum_{i=1}^N \mbox{Tr}(\widehat{g}_i (\theta)^\top \widehat{g}_i (\theta)) \\
&= \frac{1}{N} \sum_{i=1}^N || \widehat{g}_i (\theta)||^2_2
\end{align*}

In our case, $\widehat{g} (\bR) = \left[ \mbox{\textbf{vec}} [ \widehat{\bKc} - \bKc(\bR) ], \mbox{\textbf{vec}} [ \widehat{\bC} - \bC (\bR) ] \right]^\top \in \mathbb{R}^{2 d^2}$.
Assuming the associated weighting matrix is block-wise, one block for $\widehat{\bKc} - \bKc(\bR)$ and the other for $\widehat{\bC} - \bC (\bR)$,
the sum of the eigenvalues of the first block becomes $\|\widehat{\bKc} - \bKc(\bR)\|^2_2$, and $\| \widehat{\bC} - \bC (\bR) \|^2_2$ for the second.
We compute the previous terms with $\bR_1 = 0$. All together, the objective function to minimize is
\begin{align}
 \frac{1}{\| \widehat{\bKc} \|_2^2} \|\bKc(\bR) - \widehat{\bKc}\|_2^2 + \frac{1}{\| \widehat{\bC} \|_2^2} \| \bC (\bR) - \widehat{\bC} \|_2^2,
\end{align}
which equals the loss function given in \ref{eq:nphc_loss}, up to a constant.

\end{document}